\documentclass[11pt]{article}
\usepackage[
  letterpaper,
  top=1in,
  bottom=1in,
  left=1in,
  right=1in
]{geometry}
\usepackage[T1]{fontenc}
\usepackage{mathpazo}
\usepackage{amsmath,amssymb,amsthm,mathtools,bm}
\usepackage{aliascnt}
\usepackage{microtype}
\usepackage{enumitem}
\usepackage[dvipsnames]{xcolor}
\usepackage{booktabs}
\usepackage{array}
\usepackage{tabularx}
\usepackage{graphicx}
\usepackage{adjustbox}
\usepackage{float}
\usepackage{tikz}
\usepackage{subcaption}
\usepackage{wrapfig}
\usetikzlibrary{arrows.meta,backgrounds,calc,decorations.pathreplacing,fit,positioning}
\usepackage{fancyhdr}
\usepackage{appendix}
\usepackage{xurl}
\usepackage[pagebackref]{hyperref}
\usepackage[nameinlink,capitalize,noabbrev]{cleveref}

\hypersetup{
  colorlinks=true,
  urlcolor=blue,
  linkcolor=RoyalBlue,
  citecolor=OliveGreen,
  linktocpage=true,
  pdftitle={Online Algorithms Cannot Find Isolated Perceptron Solutions},
  pdfauthor={Shuangping Li and Gabe Schoenbach}
}

\definecolor{ChangeColor}{RGB}{75,0,110}

\allowdisplaybreaks
\makeatletter
\renewcommand\paragraph{\@startsection{paragraph}{4}{\z@}{1.5ex \@plus .5ex \@minus .2ex}{-1em}{\normalfont\normalsize\bfseries}}
\makeatother
\setlist[itemize]{topsep=0.35em,itemsep=0.2em,parsep=0pt,leftmargin=1.6em}
\setlist[enumerate]{topsep=0.35em,itemsep=0.25em,parsep=0pt,leftmargin=1.8em}
\newtheorem{theorem}{Theorem}[section]

\newaliascnt{proposition}{theorem}

\aliascntresetthe{proposition}

\newaliascnt{lemma}{theorem}
\newtheorem{lemma}[lemma]{Lemma}
\aliascntresetthe{lemma}

\newaliascnt{corollary}{theorem}

\aliascntresetthe{corollary}

\theoremstyle{plain}
\newaliascnt{definition}{theorem}
\newtheorem{definition}[definition]{Definition}
\aliascntresetthe{definition}

\theoremstyle{remark}
\newaliascnt{remark}{theorem}

\aliascntresetthe{remark}

\theoremstyle{plain}
\newaliascnt{claim}{theorem}

\aliascntresetthe{claim}
\numberwithin{equation}{section}

\newcommand{\R}{\mathbb R}
\providecommand{\G}{\mathcal{G}}
\providecommand{\Prob}[1]{\mathbb{P}\left(#1\right)}
\providecommand{\Ex}[1]{\mathbb{E}\left[#1\right]}
\providecommand{\Exp}[1]{\exp\left(#1\right)}
\providecommand{\A}{\mathsf{Alg}}
\providecommand{\subA}{\mathsf{SubAlg}}
\providecommand{\margin}{\kappa}
\providecommand{\solset}{\mathcal{S}(G,\margin)}
\providecommand{\isosolset}{\mathcal{S}^{h}(G,\margin)}
\providecommand{\sol}{\boldsymbol{\sigma}}

\providecommand{\threshstat}{\alpha_{\mathrm{sat}}}

\providecommand{\Gpref}{G'}
\providecommand{\Gsuff}{Y}

\providecommand{\solsuff}{z}
\providecommand{\success}{\mathsf{Success}}

\providecommand{\residual}{\theta}
\providecommand{\online}{\mathsf{ONLINE}}

\providecommand{\N}{\mathbb{N}}
\providecommand{\norm}[1]{\lVert #1\rVert}

\providecommand{\parallelfunc}{\phi}
\newcommand{\signs}{\{\pm 1\}}
\newcommand{\eps}{\varepsilon}

\newcommand{\orthant}[1]{\mathcal{O}_{#1}}
\newcommand{\Var}{\text{Var}}
\newcommand{\varW}{b_0}
\newcommand{\subgaussian}[1]{\norm{#1}_{\psi_2}}

\newcommand{\underProb}[2]{\underset{#1}{\mathbb{P}}\left(#2\right)}

\newcommand{\unique}{\mathcal{U}_{\parallelfunc, \residual}}
\newcommand{\stdnormal}{\mathcal{N}(0,I_m)}
\newcommand{\Cov}{\text{Cov}}
\newcommand{\rectangleset}{\mathcal{R}_m}
\newcommand{\score}{S_{\parallelfunc}}
\newcommand{\match}{\mathsf{Match}}
\newcommand{\nextcol}{\parallelfunc_{r}}
\newcommand{\mistake}{\mathsf{B}}

\crefname{theorem}{theorem}{theorems}
\Crefname{theorem}{Theorem}{Theorems}
\crefname{lemma}{lemma}{lemmas}
\Crefname{lemma}{Lemma}{Lemmas}
\crefname{proposition}{proposition}{propositions}
\Crefname{proposition}{Proposition}{Propositions}
\crefname{corollary}{corollary}{corollaries}
\Crefname{corollary}{Corollary}{Corollaries}
\crefname{remark}{remark}{remarks}
\Crefname{remark}{Remark}{Remarks}

\title{Online Algorithms Cannot Find Isolated Perceptron Solutions}
\author{Shuangping Li\thanks{Department of Statistics and Data Science, Yale University. \texttt{shuangping.li@yale.edu}.}
\and
Gabe Schoenbach\thanks{Department of Computer Science, University of Chicago. \texttt{gschoenbach@uchicago.edu}.}}
\date{\today}

\begin{document}
\hypersetup{pageanchor=false}
\maketitle

\begin{abstract}
The binary perceptron problem asks for a length-$n$ sign vector in the intersection of $\lfloor \alpha n\rfloor$-many independent random Gaussian half-spaces. A striking feature of this model is strong freezing: at every constraint density $\alpha > 0$, all but a vanishing fraction of solutions are \emph{isolated} --- $\Omega(n)$-far from all others, in Hamming distance. Such solutions are believed to be algorithmically inaccessible. Nevertheless, efficient algorithms find solutions at some constraint densities $\alpha$. One possible resolution of this apparent contradiction is that the solution space also contains rare well-connected structures, and that algorithms succeed by locating them. Gong, Huang, Li and Sellke (FOCS 2026)~\cite{GongHuangLiSellke2026} proved this for stable algorithms, by showing that such algorithms cannot locate isolated solutions with probability more than $\approx 0.84$.

In this work, we study online algorithms, which receive the columns of the constraint matrix one by one and sign each irrevocably upon arrival. This is a widely used class of algorithms in discrepancy minimization, and it includes the algorithm of Jo and Lee (2026)~\cite{JL26}, which has one of the best provable guarantees for the binary perceptron. We prove that at any constraint density, any online algorithm outputs an isolated solution with probability $o(1)$. 

Our proof is based on the observation that the signs an online algorithm has already fixed can be completed to at most one isolated solution, so a successful algorithm must choose the unique correct sign for each of the last few columns. Using the exchangeability of these columns, we show that such an algorithm must match a \emph{suffix-parallel baseline}, which signs each of them as if it were the next column to arrive. We then use an anti-concentration argument to show that such a baseline rarely produces a unique completion.
\end{abstract}

\thispagestyle{empty}
\newpage
\begingroup
\small
\tableofcontents
\endgroup
\thispagestyle{empty}
\newpage
\hypersetup{pageanchor=true}
\setcounter{page}{1}

\section{Introduction}\label{sec:introduction}

Random constraint satisfaction problems are a central testing ground for average-case complexity. For many such problems, efficient algorithms are only known to find solutions in a restricted parameter range, even though solutions exist with high probability over a wider range. A major goal of theoretical computer science is to understand the source of these \emph{statistical-to-computational} gaps. A recurring theme has been the suggestion that the relative position of solutions --- in other words, the \emph{geometry} of the solution space --- might guide the success of efficient algorithms~\cite{AC08,Gamarnik2021}. In this paper we study the binary perceptron problem, a random constraint satisfaction problem whose solution space has a particularly striking geometry.

In the binary perceptron problem, we seek a point of the hypercube $\signs^n$ that lies in the intersection of $m$ random half-spaces. The model goes back to the study of neural networks with binary weights~\cite{Ros58,Cover1965,GardnerDerrida1988,KrauthMezard1989}, and it remains a fundamental model for understanding learning with discrete weights~\cite{BBCILSZ16}. In theoretical computer science, it is closely related to discrepancy minimization~\cite{Spe85,Ban10,LM15}. In fact, its symmetric version is exactly the problem of finding a low-discrepancy signing of a random matrix~\cite{AubinPerkinsZdeborova2019}. The model also has close connections to statistical physics~\cite{Gardner1988,HWK13,HuangKabashima2014}, probability theory~\cite{Tal11,DingSun2019,NakajimaSun2023,Hua24}, high-dimensional geometry~\cite{KimRoche1998,Tal99,Tal00}, and, more recently, lattice problems and cryptography~\cite{VV25,BBMPRSZ26}. The perceptron has also become a testing ground for questions about algorithms and computational hardness~\cite{AbbeLiSly2022RareCluster,GamarnikKizildagPerkinsXu2022,LiSchrammZhou2025,GongHuangLiSellke2026,HSS26,JL26}.

\subsection{The binary perceptron problem}\label{sec:background}

Given an $m \times n$ matrix $G$ with i.i.d.\ standard Gaussian entries, the binary perceptron problem asks for a sign vector $\sigma \in \signs^n$ such that $G\sigma \ge \margin\sqrt{n}$ entrywise (see \Cref{def:perceptron}). This is a constraint satisfaction problem with $m$ constraints: one linear inequality for each row of $G$. A vector $\sigma$ satisfying all constraints is called a solution. The fixed parameter $\margin \in \R$ is called the margin.\footnote{
    Many other perceptron models are obtained by requiring instead that $G\sigma/\sqrt{n} \in K$ entrywise, for a fixed set $K \subseteq \R$~\cite{GongLi2026}. The binary perceptron corresponds to $K = [\margin, \infty)$, and $K = [-\margin, \margin]$ gives the symmetric binary perceptron ($\mathsf{SBP}$) of Aubin, Perkins and Zdeborov\'a~\cite{AubinPerkinsZdeborova2019}.
} We are interested in the proportional regime $m = \lfloor \alpha n \rfloor$, where the constraint density $\alpha > 0$ (the ratio of the number of constraints to the dimension) is fixed as $n \to \infty$.

The first natural question about the binary perceptron is for which $\alpha$ do solutions exist. In the late 1980s, physicists including Krauth and M\'ezard~\cite{KrauthMezard1989} conjectured, based on the non-rigorous replica method, that the binary perceptron has a sharp satisfiability phase transition at a specific constraint density $\threshstat(\margin)$.\footnote{
    In other words, solutions exist with high probability whenever $\alpha < \threshstat(\margin)$, but whenever $\alpha > \threshstat(\margin)$, with high probability there are no solutions.
} At $\margin = 0$, this conjecture was later confirmed by a long line of work (see, e.g.,~\cite{DingSun2019,Hua24}, and \Cref{sec:related-work} for more).

Other natural questions are algorithmic: at which constraint densities can a solution be found efficiently, and what kind of solutions do efficient algorithms find? For many random constraint satisfaction problems, the answer is closely tied to the geometry of the solution space~\cite{KMRTSZ07,AC08,Gamarnik2021}. As such, we now turn to the geometry of the perceptron, which differs sharply from that of most other random models.

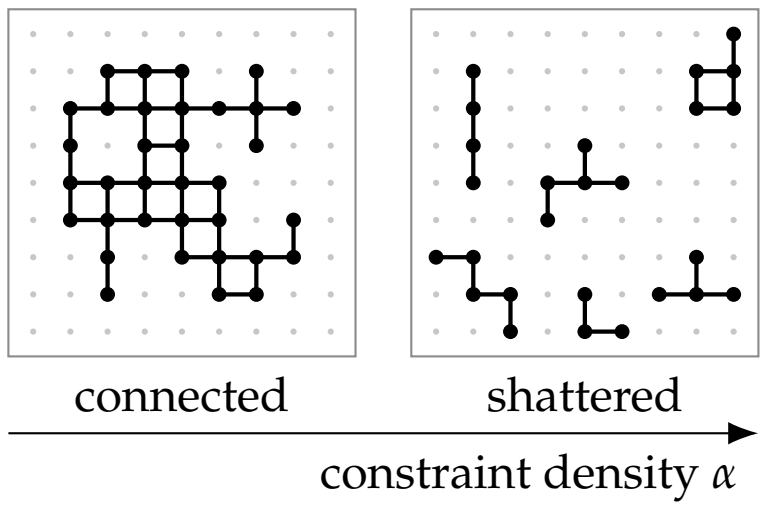
\begin{wrapfigure}{R}{0.34\textwidth}
    \centering
    \vspace{-0.5\baselineskip}
    \begin{tikzpicture}[scale=0.82, every node/.style={font=\footnotesize}]
        \begin{scope}[xshift=0.00cm]
            \draw[black!45] (-0.20,-0.20) rectangle (2.60,2.60);
            \foreach \x in {0,...,8} \foreach \y in {0,...,8} \fill[black!22] (0.3*\x,0.3*\y) circle (0.025);
            \draw[black, thick] (0.30,0.90) -- (0.30,1.20) (0.30,0.90) -- (0.60,0.90) (0.30,1.20) -- (0.30,1.50) (0.30,1.20) -- (0.60,1.20) (0.30,1.50) -- (0.30,1.80) (0.30,1.80) -- (0.60,1.80) (0.60,0.30) -- (0.60,0.60) (0.60,0.60) -- (0.60,0.90) (0.60,0.90) -- (0.60,1.20) (0.60,0.90) -- (0.90,0.90) (0.60,1.20) -- (0.90,1.20) (0.60,1.80) -- (0.60,2.10) (0.60,1.80) -- (0.90,1.80) (0.60,2.10) -- (0.90,2.10) (0.90,0.90) -- (0.90,1.20) (0.90,0.90) -- (1.20,0.90) (0.90,1.20) -- (0.90,1.50) (0.90,1.20) -- (1.20,1.20) (0.90,1.50) -- (0.90,1.80) (0.90,1.50) -- (1.20,1.50) (0.90,1.80) -- (0.90,2.10) (0.90,1.80) -- (1.20,1.80) (0.90,2.10) -- (1.20,2.10) (1.20,0.60) -- (1.20,0.90) (1.20,0.60) -- (1.50,0.60) (1.20,0.90) -- (1.20,1.20) (1.20,0.90) -- (1.50,0.90) (1.20,1.20) -- (1.20,1.50) (1.20,1.20) -- (1.50,1.20) (1.20,1.50) -- (1.20,1.80) (1.20,1.80) -- (1.20,2.10) (1.20,1.80) -- (1.50,1.80) (1.50,0.30) -- (1.50,0.60) (1.50,0.30) -- (1.80,0.30) (1.50,0.60) -- (1.50,0.90) (1.50,0.60) -- (1.80,0.60) (1.50,0.90) -- (1.50,1.20) (1.50,1.80) -- (1.80,1.80) (1.80,0.30) -- (1.80,0.60) (1.80,0.60) -- (2.10,0.60) (1.80,1.50) -- (1.80,1.80) (1.80,1.80) -- (1.80,2.10) (1.80,1.80) -- (2.10,1.80) (2.10,0.60) -- (2.10,0.90);
            \foreach \x/\y in {0.30/0.90, 0.30/1.20, 0.30/1.50, 0.30/1.80, 0.60/0.30, 0.60/0.60, 0.60/0.90, 0.60/1.20, 0.60/1.80, 0.60/2.10, 0.90/0.90, 0.90/1.20, 0.90/1.50, 0.90/1.80, 0.90/2.10, 1.20/0.60, 1.20/0.90, 1.20/1.20, 1.20/1.50, 1.20/1.80, 1.20/2.10, 1.50/0.30, 1.50/0.60, 1.50/0.90, 1.50/1.20, 1.50/1.80, 1.80/0.30, 1.80/0.60, 1.80/1.50, 1.80/1.80, 1.80/2.10, 2.10/0.60, 2.10/0.90, 2.10/1.80} \fill[black] (\x,\y) circle (0.06);
            \node[below] at (1.20,-0.20) {connected};
        \end{scope}
        \begin{scope}[xshift=3.25cm]
            \draw[black!45] (-0.20,-0.20) rectangle (2.60,2.60);
            \foreach \x in {0,...,8} \foreach \y in {0,...,8} \fill[black!22] (0.3*\x,0.3*\y) circle (0.025);
            \draw[black, thick] (0.00,0.60) -- (0.30,0.60) (0.30,0.30) -- (0.30,0.60) (0.30,0.30) -- (0.60,0.30) (0.30,1.20) -- (0.30,1.50) (0.30,1.50) -- (0.30,1.80) (0.30,1.80) -- (0.30,2.10) (0.60,0.00) -- (0.60,0.30) (0.90,0.90) -- (0.90,1.20) (0.90,1.20) -- (1.20,1.20) (1.20,0.00) -- (1.20,0.30) (1.20,0.00) -- (1.50,0.00) (1.20,1.20) -- (1.20,1.50) (1.20,1.20) -- (1.50,1.20) (1.80,0.30) -- (2.10,0.30) (2.10,0.30) -- (2.10,0.60) (2.10,0.30) -- (2.40,0.30) (2.10,1.80) -- (2.10,2.10) (2.10,1.80) -- (2.40,1.80) (2.10,2.10) -- (2.40,2.10) (2.40,1.80) -- (2.40,2.10) (2.40,2.10) -- (2.40,2.40);
            \foreach \x/\y in {0.00/0.60, 0.30/0.30, 0.30/0.60, 0.30/1.20, 0.30/1.50, 0.30/1.80, 0.30/2.10, 0.60/0.00, 0.60/0.30, 0.90/0.90, 0.90/1.20, 1.20/0.00, 1.20/0.30, 1.20/1.20, 1.20/1.50, 1.50/0.00, 1.50/1.20, 1.80/0.30, 2.10/0.30, 2.10/0.60, 2.10/1.80, 2.10/2.10, 2.40/0.30, 2.40/1.80, 2.40/2.10, 2.40/2.40} \fill[black] (\x,\y) circle (0.06);
            \node[below] at (1.20,-0.20) {shattered};
        \end{scope}
        \draw[-{Latex[length=2mm]}] (-0.20,-0.82) -- (5.85,-0.82) node[below left] {constraint density $\alpha$};
    \end{tikzpicture}
    \vspace{-0.6\baselineskip}
    \caption{Phase diagram of a random constraint satisfaction problem such as random $k$-SAT. Black dots are solutions.}
    \label{fig:shattering}
\end{wrapfigure}
In most random constraint satisfaction problems, such as random $k$-SAT or the coloring of random graphs, the solutions start to shatter into many well-separated clusters\footnote{Connect two solutions by an edge if their Hamming distance is small. The clusters are the connected components of the resulting graph on the set of solutions.} only above some positive constraint density~\cite{AchlioptasRicciTersenghi2006,KMRTSZ07,AC08,Mol18} (see \Cref{fig:shattering}). Surprisingly, the set of solutions of the binary perceptron is conjectured to shatter at every positive constraint density, and in the most extreme way: almost every solution is \emph{isolated}, forming a cluster by itself, with no other solution within linear Hamming distance of it~\cite{KrauthMezard1989,HWK13,HuangKabashima2014}. This property --- known as \emph{strong freezing} --- was first proved for the symmetric binary perceptron~\cite{PerkinsXu2021,AbbeLiSly2022Contiguity}, and later for a general class of perceptron models~\cite{GongLi2026}. In particular, it holds for the binary perceptron at every margin $\margin$ and every constraint density $\alpha > 0$.

At first glance, the presence of strong freezing at every constraint density suggests that perceptron solutions should always be hard to find, since there is no apparent way for an algorithm to make partial progress towards a solution with no other solution nearby~\cite{ZM08b}. Indeed, in many random problems, the onset of shattering is believed to coincide with the failure of efficient algorithms~\cite{KMRTSZ07,AC08,ZK07}. Nevertheless, there are some positive constraint densities at which efficient algorithms provably do find perceptron solutions~\cite{KimRoche1998,AbbeLiSly2022RareCluster,LiSchrammZhou2025,JL26}. For example, at $\margin = 0$, the online algorithm of Jo and Lee~\cite{JL26}, which signs the columns of $G$ one at a time, finds a solution with high probability for every $\alpha < 0.327$.

\begin{figure}[!htb]
    \centering
    \begin{tikzpicture}[every node/.style={font=\small}]
        \draw[rounded corners=12pt, black!45] (-0.40,-0.40) rectangle (10.30,4.90);
        \foreach \x in {0,...,33} \foreach \y in {0,...,15} \fill[black!22] (0.3*\x,0.3*\y) circle (0.025);
        \draw[blue!60!black, thick] (0.90,0.60) -- (1.20,0.60) (1.20,0.60) -- (1.20,0.90) (1.20,0.90) -- (1.20,1.20) (1.20,1.20) -- (1.50,1.20) (1.50,1.20) -- (1.80,1.20) (1.80,1.20) -- (1.80,1.50) (1.80,1.50) -- (1.80,1.80) (1.80,1.80) -- (2.10,1.80) (2.10,1.80) -- (2.40,1.80) (2.40,1.80) -- (2.70,1.80) (2.70,1.80) -- (3.00,1.80) (3.00,1.80) -- (3.30,1.80) (3.30,1.50) -- (3.30,1.80) (3.30,1.80) -- (3.60,1.80) (3.60,1.80) -- (3.60,2.10) (3.60,1.80) -- (3.90,1.80) (3.60,2.10) -- (3.90,2.10) (3.90,1.80) -- (3.90,2.10) (3.90,2.10) -- (3.90,2.40) (3.90,2.10) -- (4.20,2.10) (4.20,2.10) -- (4.50,2.10) (4.50,2.10) -- (4.80,2.10) (4.80,1.80) -- (4.80,2.10) (4.80,1.80) -- (5.10,1.80) (5.10,1.20) -- (5.10,1.50) (5.10,1.50) -- (5.10,1.80) (5.10,1.80) -- (5.40,1.80) (5.40,1.80) -- (5.70,1.80) (5.70,1.80) -- (6.00,1.80) (6.00,1.20) -- (6.00,1.50) (6.00,1.50) -- (6.00,1.80) (6.00,1.50) -- (6.30,1.50) (6.30,1.50) -- (6.60,1.50) (6.60,1.50) -- (6.60,1.80) (6.60,1.50) -- (6.90,1.50) (6.60,1.80) -- (6.90,1.80) (6.90,1.50) -- (6.90,1.80) (6.90,1.50) -- (7.20,1.50) (7.20,1.50) -- (7.50,1.50) (7.50,1.50) -- (7.80,1.50) (7.80,1.50) -- (8.10,1.50) (8.10,1.50) -- (8.40,1.50) (8.40,0.90) -- (8.40,1.20) (8.40,0.90) -- (8.70,0.90) (8.40,1.20) -- (8.40,1.50) (8.40,1.20) -- (8.70,1.20) (8.40,1.50) -- (8.70,1.50) (8.70,0.90) -- (8.70,1.20) (8.70,1.20) -- (8.70,1.50);
        \foreach \x/\y in {0.90/0.60, 1.20/0.60, 1.20/0.90, 1.20/1.20, 1.50/1.20, 1.80/1.20, 1.80/1.50, 1.80/1.80, 2.10/1.80, 2.40/1.80, 2.70/1.80, 3.00/1.80, 3.30/1.50, 3.30/1.80, 3.60/1.80, 3.60/2.10, 3.90/1.80, 3.90/2.10, 3.90/2.40, 4.20/2.10, 4.50/2.10, 4.80/1.80, 4.80/2.10, 5.10/1.20, 5.10/1.50, 5.10/1.80, 5.40/1.80, 5.70/1.80, 6.00/1.20, 6.00/1.50, 6.00/1.80, 6.30/1.50, 6.60/1.50, 6.60/1.80, 6.90/1.50, 6.90/1.80, 7.20/1.50, 7.50/1.50, 7.80/1.50, 8.10/1.50, 8.40/0.90, 8.40/1.20, 8.40/1.50, 8.70/0.90, 8.70/1.20, 8.70/1.50} \fill[blue!60!black] (\x,\y) circle (0.065);
        \foreach \x/\y in {0.00/0.30, 0.00/1.80, 0.00/3.00, 0.00/4.50, 0.30/1.20, 0.60/0.00, 0.60/2.10, 0.60/3.90, 1.20/2.70, 1.50/4.50, 1.80/0.30, 1.80/3.30, 2.10/4.20, 2.40/0.00, 2.40/0.90, 3.00/0.30, 3.00/2.70, 3.00/3.90, 3.30/4.50, 3.60/3.30, 3.90/0.30, 3.90/3.90, 4.20/0.90, 4.20/3.00, 4.20/4.50, 4.50/0.00, 4.80/0.60, 4.80/3.30, 5.10/2.70, 5.10/3.90, 5.70/0.00, 5.70/3.00, 5.70/4.50, 6.00/3.60, 6.30/0.60, 6.30/2.70, 6.30/4.20, 6.60/0.00, 6.60/3.30, 6.90/3.90, 7.20/0.60, 7.20/3.00, 7.20/4.50, 7.50/0.00, 7.80/2.40, 7.80/3.30, 8.10/0.30, 8.10/4.50, 8.40/2.70, 8.70/3.90, 9.00/0.00, 9.00/2.10, 9.00/3.00, 9.30/0.60, 9.30/4.50, 9.60/3.30, 9.90/0.00, 9.90/1.50, 9.90/2.40, 9.90/4.20} \fill[red!70!black] (\x,\y) circle (0.075);
        \draw[dashed, red!70!black] (2.70,3.30) -- (1.80,4.20) -- (0.90,3.30) -- (1.80,2.40) -- cycle;
        \draw[red!70!black, -{Latex[length=1.5mm]}] (1.80,3.30) -- (2.70,3.30) node[midway, above] {$\delta n$};
        \fill[red!70!black] (1.6,-0.9) circle (0.075);
        \node[red!70!black, right] at (1.7,-0.9) {isolated solutions};
        \draw[blue!60!black, thick] (5.3,-0.9) -- (5.6,-0.9); \fill[blue!60!black] (5.3,-0.9) circle (0.065) (5.6,-0.9) circle (0.065);
        \node[blue!60!black, right] at (5.7,-0.9) {rare well-connected cluster};
        \node[align=center] (alg) at (11.60,3.90) {efficient\\ algorithms};
        \draw[-{Latex[length=2mm]}, thick] (alg.south west) -- (8.82,1.55);
    \end{tikzpicture}
    \caption{Almost all solutions are isolated (red), with no other solution within Hamming distance $\delta n$ (dashed). An exponentially small fraction of the solutions form a well-connected cluster (blue) of very large diameter, and efficient algorithms are believed to find solutions in such clusters. Two solutions are joined by an edge if their Hamming distance is one.}
    \label{fig:clusters}
\end{figure}
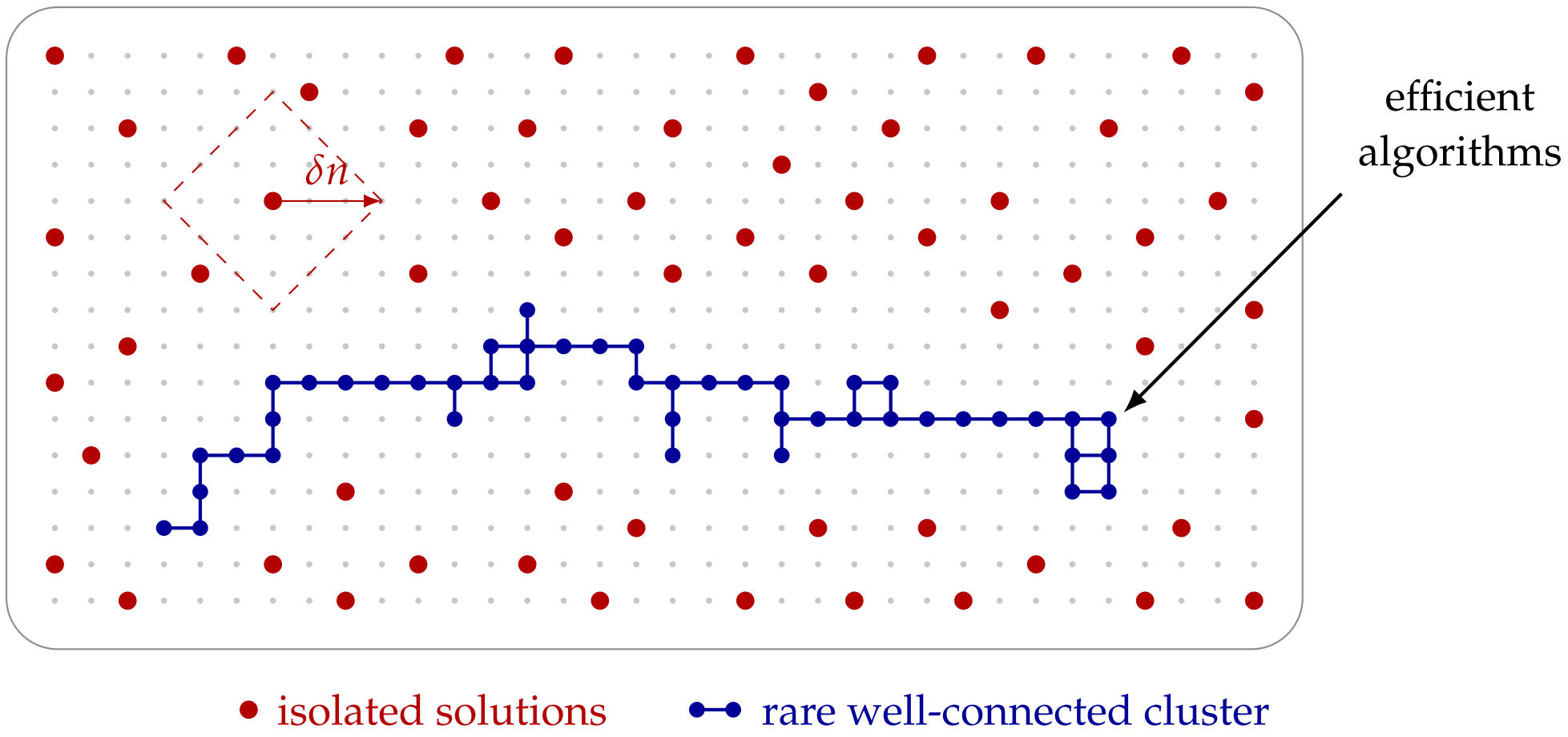

To address this seemingly contradictory phenomenon,~\cite{BaldassiEtAl2015,BaldassiEtAl2020} proposed the following picture (based on non-rigorous methods from statistical physics). Although almost all solutions are isolated, an exponentially-small fraction of them may belong to well-connected clusters with large diameter, and efficient algorithms may succeed by finding these rare clusters (see \Cref{fig:clusters}). To make this picture rigorous, one needs two facts. First, such rare clusters exist. This is by now well supported: Abbe, Li and Sly~\cite{AbbeLiSly2022RareCluster} proved that at small constraint densities the set of solutions contains a rare, well-connected cluster of nearly maximal diameter, and further evidence for such clusters comes from~\cite{HSS26,BAKZ24,Bar24,Bar25,Bar26,Sto25}. Second, the isolated solutions themselves are hard to find. This second point was raised by El Alaoui and Gamarnik~\cite{AG25} as an open problem in their work on the hardness of sampling, and was implicit in the strong freezing picture of Perkins and Xu~\cite{PerkinsXu2021}. This question is the subject of our work.

\subsection{Main result}\label{sec:main-result}

\Cref{thm:informal} addresses the second of the two facts needed for the picture of ~\cite{BaldassiEtAl2015,BaldassiEtAl2020}, namely that isolated solutions are hard (for online algorithms) to find. Combined with strong freezing, it shows that online algorithms can only find atypical solutions. The same holds for the symmetric binary perceptron, where online discrepancy minimization algorithms~\cite{BansalSpencer2020,JL26b} find solutions at positive constraint densities and strong freezing is also known~\cite{PerkinsXu2021,AbbeLiSly2022Contiguity}.

\begin{theorem}[Informal version of \Cref{thm:main,thm:general-intro}]\label{thm:informal}
    At every constraint density and every margin, every online algorithm finds an isolated solution of the binary perceptron with probability $o(1)$. The same holds for the symmetric binary perceptron and for many other perceptron models.
\end{theorem}

Online algorithms are those which receive the columns of $G$ one at a time and must choose the sign of each column irrevocably when it arrives (see \Cref{def:online}).\footnote{
    Other than this requirement, the algorithm is completely unrestricted. It may be randomized, remember everything it has seen, and run for an unbounded amount of time.
} This is a natural class of algorithms for the perceptron problem; it includes the algorithm of~\cite{JL26}, which has the best provable\footnote{
    Huang, Sellke and Sun~\cite{HSS26} recently constructed an approximate message passing algorithm that is optimal among Lipschitz algorithms, and whose threshold is given by a one-dimensional stochastic control problem. This algorithm is expected to succeed at higher constraint densities than the algorithm of~\cite{JL26}. Confirming this for the binary perceptron requires further work, including a numerical solution of the resulting variational problem.
} guarantee at zero margin. Online algorithms are also widely used in discrepancy minimization, where the vectors arrive one at a time and each must be signed on arrival so that the signed sum stays small~\cite{BansalSpencer2020,BJSS20,ALS21b,KRR24}.

The first hardness result for finding isolated solutions is due to Gong, Huang, Li and Sellke~\cite{GongHuangLiSellke2026}. They showed that a stable algorithm, whose output changes only slightly when the input is slightly resampled, finds a linearly isolated solution with probability at most roughly $0.84233$. Since the bound of~\cite{GongHuangLiSellke2026} is a constant, it does not rule out, for example, a stable algorithm that finds an isolated solution with probability $0.8$ --- see \Cref{sec:related-work} for further discussion. Moreover, it is not clear whether the online algorithm of~\cite{JL26} is stable in the sense of~\cite{GongHuangLiSellke2026}, so their results do not tell us whether this algorithm can find isolated solutions. 

In contrast with the constant bound of~\cite{GongHuangLiSellke2026}, the success probability in \Cref{thm:informal} vanishes as $n \to \infty$, so our result rules out online algorithms that find isolated solutions with any constant probability. We highlight three further features of our result. First, it applies to a much weaker notion of isolation: in fact, it suffices that no other solution lies within Hamming distance $\log^{4 + \eps}(n)$ of the output for any $\eps > 0$, instead of linear distance (see \Cref{thm:main}). Second, our argument is robust to the form of the constraints: it extends to the symmetric binary perceptron and to many other perceptron models (\Cref{thm:general-intro}). Third, our proof technique is quite different from that of~\cite{GongHuangLiSellke2026} and from the many hardness results for online algorithms that are based on the overlap gap property. We give an overview of the argument in \Cref{sec:technical-overview}.

\subsection{Related work}\label{sec:related-work}

\paragraph{Satisfiability threshold.} Using the non-rigorous replica method, Krauth and M\'ezard~\cite{KrauthMezard1989} conjectured an explicit value for the satisfiability threshold $\threshstat(\margin)$, for instance $\threshstat(0) \approx 0.833$. Unlike the spherical perceptron~\cite{ST03,Sto13a}, the binary perceptron is nonconvex, and this conjecture remained open for decades. At $\margin = 0$, Ding and Sun~\cite{DingSun2019} proved that solutions exist below this value, and Huang~\cite{Hua24} proved that there are none above it, both assuming numerical conditions that were later verified~\cite{Nak25,Shm26}. Other results on the threshold and the number of solutions include~\cite{KimRoche1998,Tal99,Tal00,Tal11,BNSX22,Xu2021,NakajimaSun2023,AT24}.

\paragraph{Algorithms.} Several efficient algorithms provably find solutions at positive constraint densities, including a multistage majority algorithm~\cite{KimRoche1998} and an algorithm based on discrepancy theory~\cite{LiSchrammZhou2025}. Jo and Lee~\cite{JL26} determined the exact threshold $\alpha_{\mathrm{on}}(\margin)$ for online algorithms, with $0.327 < \alpha_{\mathrm{on}}(0) < 0.367$, still well below $\threshstat(0)$. Huang, Sellke and Sun~\cite{HSS26} characterized the performance of Lipschitz algorithms for a broad class of perceptron models. Message-passing heuristics have also been studied extensively~\cite{BraunsteinZecchina2006,BBBZ07,Bal09,BB15}.

\paragraph{Overlap gap property.} The most widely used approach to show algorithmic hardness for random search problems is the overlap gap property (OGP), introduced by Gamarnik and Sudan~\cite{GS17a} (see~\cite{Gamarnik2021} for a survey). The OGP says that the overlaps between near-optimal solutions avoid an intermediate range, which rules out stable algorithms, since their outputs change gradually as the input is resampled~\cite{CGPR19,GamarnikJagannathWein2020}. Variants involving many solutions give sharp hardness results for independent sets~\cite{RV17,Wei22}, random $k$-SAT~\cite{BH22} and perceptron models~\cite{GamarnikKizildagPerkinsXu2022,LiSchrammZhou2025,HSS26}. OGP-based arguments also rule out sequential and online algorithms~\cite{GS17b,GamarnikKizildagPerkinsXu2023,GKW25}. Related results include~\cite{HS25b,HS25a,KL26,DGH25,DhawanKizildagMaitra2025,BhamidiGamarnikGong2026,GRS26,BBMPRSZ26}. However, an OGP does not necessarily imply algorithmic hardness~\cite{LS25}.

An OGP for isolated solutions also seems difficult to establish. Conditioning on isolation creates complicated dependencies between the constraints, and since there are exponentially many isolated solutions, a first moment argument would need sharp large deviation estimates.

\paragraph{Hardness of finding isolated perceptron solutions.} The constant in the bound of Gong, Huang, Li and Sellke~\cite{GongHuangLiSellke2026} appears to be intrinsic to their approach. It comes from balancing a lower bound derived from Pitt's correlation inequality~\cite{Pitt1982} against a trivial upper bound, so the argument cannot be pushed to an $o(1)$ bound simply by optimizing its parameters. One might also hope to boost a stable algorithm that succeeds with constant probability (for instance by running it several times) into an algorithm that finds an isolated solution with probability close to $1$. We do not expect this to work. Even if one of several runs finds an isolated solution, there is not an obvious efficient way to tell which one succeeded, since checking that a solution is isolated requires ruling out all other solutions in a Hamming ball of linear radius.

\paragraph{Symmetric binary perceptron.} For the symmetric model, the satisfiability threshold was established in~\cite{AubinPerkinsZdeborova2019,PerkinsXu2021,AbbeLiSly2022Contiguity}, strong freezing was proved in~\cite{PerkinsXu2021,AbbeLiSly2022Contiguity}, and El Alaoui and Gamarnik~\cite{AG25} proved that stable algorithms cannot approximately sample its solutions. Further results on its solution space and its algorithmic hardness include~\cite{Alt23,SS23,BaldassiEtAl2020,GamarnikKizildagPerkinsXu2022,GamarnikKizildagPerkinsXu2023,VV25}.

\paragraph{Discrepancy minimization.} The perceptron has many connections to discrepancy minimization, which, given vectors $v_1, \dots, v_n \in \R^m$, asks for signs $\sigma \in \signs^n$ that make every coordinate of $\sum_t \sigma_t v_t$ small (see~\cite{Mat99} for background). The symmetric binary perceptron is the average-case version of this problem, with the columns of $G$ as the vectors $v_t$. Therefore algorithms for discrepancy minimization give algorithms for the symmetric binary perceptron, which find its solutions when the margin is large (see, e.g.,~\cite{Ban10,LM15,Rot17,ES18}). Discrepancy minimization algorithms can also find solutions of the binary perceptron~\cite{LiSchrammZhou2025}.

\paragraph{Online discrepancy.} In the online version of discrepancy minimization, the vectors arrive one at a time. Fiedler, Jackson, Lacker and Niles-Weed~\cite{FJLN26} identified the optimal online value for Gaussian vectors in the proportional regime, and Jo and Lee~\cite{JL26b} showed that it is a universal sharp threshold, which gives the exact online threshold for the symmetric binary perceptron (see also~\cite{BansalSpencer2020,BJSS20,AT26,ALS21b,KRR24,Ade26}). Applied to the columns of $G$, all of these vector balancing algorithms are online algorithms in the sense of \Cref{def:online}, so our results apply to them. Similar questions arise for matrix discrepancy, where the vectors are replaced by random symmetric matrices and the goal is to make the operator norm of the signed sum small. There, too, there is a gap between the existence of solutions and the success of stable and online algorithms~\cite{KL26}, see also~\cite{KOWZ24}.

\subsection{Roadmap}

\Cref{sec:preliminaries} collects notation and definitions. In \Cref{sec:technical-overview} we state our main results and give a technical overview of our proofs, including the proof of \Cref{thm:main} in \Cref{sec:all-together}. In \Cref{sec:proofs} we give the complete proofs enabling our main results, including the proof of \Cref{thm:general-intro} in \Cref{sec:general-perceptrons}. The rest of our technical results (additional probabilistic facts, technical lemmas, and a reduction to deterministic algorithms) are in Appendix~\ref{sec:additional-proofs}.

\section{Preliminaries}\label{sec:preliminaries}

\subsection{Notation}
We denote by $\stdnormal$ the standard Gaussian distribution on $\R^m$, that is, the distribution of a vector with i.i.d.\ $\mathcal{N}(0,1)$ entries. We write $\G^{m,n}$ for the distribution of an $m \times n$ matrix with i.i.d.\ $\mathcal{N}(0,1)$ entries, so the columns of $G \sim \G^{m,n}$ are i.i.d.\ with distribution $\stdnormal$. For any string $x$, we let $x_{i}$ denote the $i^{\text{th}}$ entry of $x$ and $x_{1:i}$ denote the first $i$ entries of $x$. We let $\|$ denote string or column-vector concatenation. For any random variable $X$, its subgaussian norm is $\norm{X}_{\psi_2} := \inf \left\{c > 0: \Ex{\Exp{\frac{X^2}{c^2}}} \leq 2\right\}$. For any $u \in \R^m$, the \emph{upper orthant $\orthant{u} \subseteq \R^m$} is $\orthant{u} := \{x \in \R^m: x_j \geq u_j \; \forall j \in [m]\}.$ We denote by $E^c$ the complement of any event $E$. We say that an event holds with high probability if its probability tends to $1$ as $n \to \infty$. For measurability, we always work with the Borel $\sigma$-algebra, which on a finite set such as $\signs^n$ is the discrete $\sigma$-algebra. This ensures that the output of an online algorithm is a random variable, so that its success probability is well defined.

\subsection{Definitions}

\begin{definition}[Binary perceptron]\label{def:perceptron}
    An instance of the binary perceptron problem is an $m \times n$ matrix $G \sim \G^{m,n}$ together with a margin $\margin \in \R$. The set of solutions to the instance $(G, \margin)$ is
    \begin{align}
        \solset = \bigcap_{j = 1}^m \left\{\sol \in \signs^n: \langle G_j, \sol \rangle \geq \margin \sqrt{n}\right\},
    \end{align}
    where $\{G_j\}_{j \in [m]}$ are the rows of $G$. More generally, for a set $K \subseteq \R$, the corresponding set of solutions is
    \begin{align}
        \mathcal{S}(G, K) = \bigcap_{j = 1}^m \left\{\sol \in \signs^n: \langle G_j, \sol \rangle/\sqrt{n} \in K\right\},
    \end{align}
    so that $\solset = \mathcal{S}(G, [\margin, \infty))$, and $K = [-\margin, \margin]$ gives the symmetric binary perceptron ($\mathsf{SBP}$).
\end{definition}

\begin{definition}[Isolated solutions]
    The Hamming distance between $\sigma, \tau \in \signs^n$ is $d(\sigma, \tau) = |\{i \in [n] : \sigma_i \neq \tau_i\}|$, and the Hamming ball of radius $r$ around $\sigma$ is denoted by $B_r(\sigma) = \{\tau \in \signs^n: d(\sigma, \tau) \leq r\}$. For any $h = h(n)$, a solution $\sol \in \solset$ is \emph{$h$-isolated} if
    \begin{align}
        \solset \cap B_{h}(\sol) = \{\sol\}.
    \end{align}
    We let $\isosolset$ denote the set of $h$-isolated solutions. We define the set $\mathcal{S}^h(G, K)$ of $h$-isolated solutions of $\mathcal{S}(G, K)$ in the same way.
\end{definition}

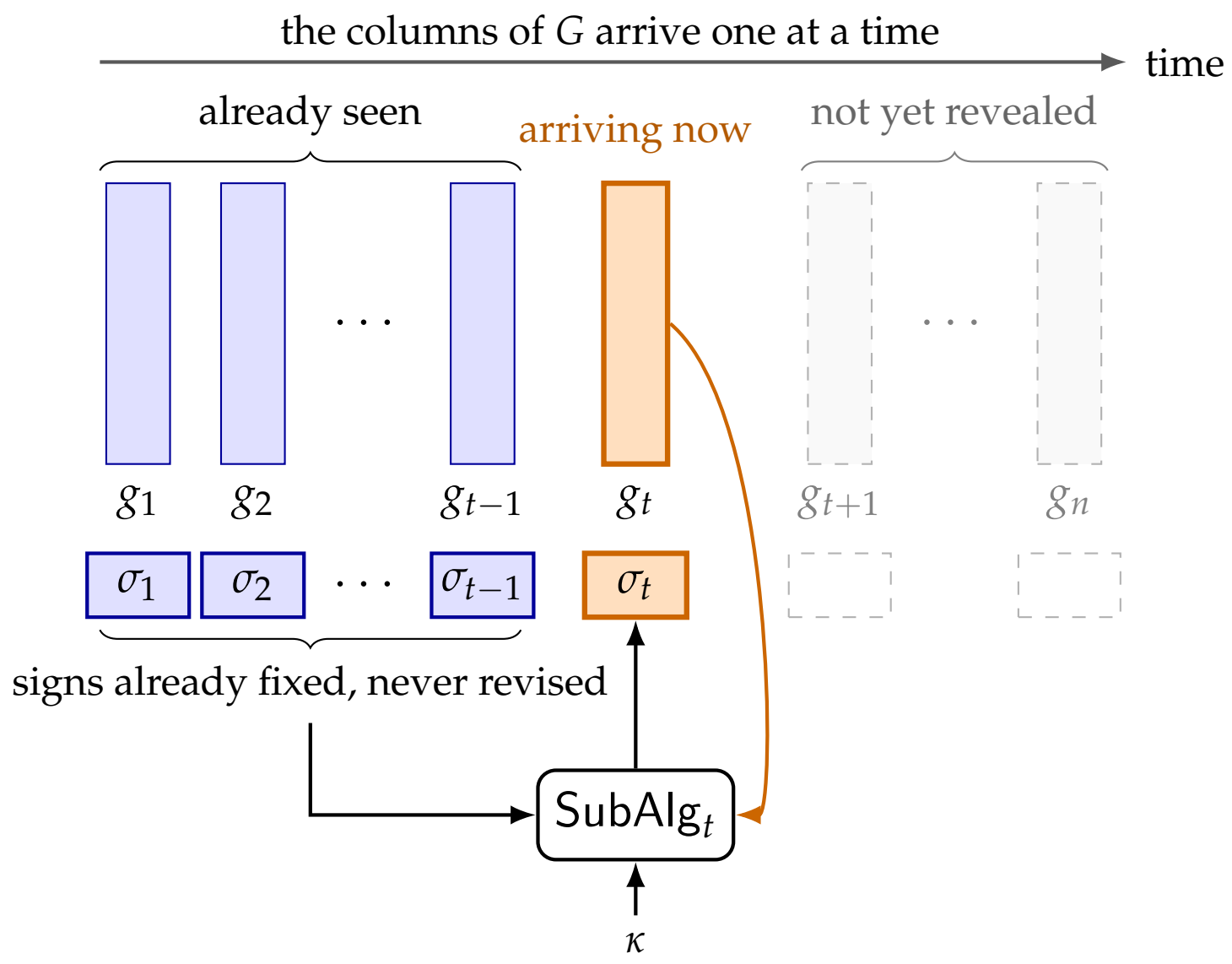
\begin{figure}[t]
    \centering
    \begin{tikzpicture}[
        >={Latex[length=2mm]},
        col/.style={minimum width=0.5cm, minimum height=2.2cm, inner sep=0pt},
        sgn/.style={minimum width=0.8cm, minimum height=0.5cm, inner sep=0pt, font=\small},
        lab/.style={font=\footnotesize}]
        \foreach \x/\i in {0/1, 0.9/2, 2.7/{t-1}} {
            \node[col, draw=blue!60!black, fill=blue!12] at (\x,1.1) {};
            \node[font=\small] at (\x,-0.3) {$g_{\i}$};
            \node[sgn, draw=blue!60!black, fill=blue!12, thick] at (\x,-0.95) {$\sigma_{\i}$};
        }
        \node at (1.8,1.1) {$\cdots$};
        \node at (1.8,-0.95) {$\cdots$};
        \node[col, draw=orange!80!black, fill=orange!25, very thick] (cur) at (3.9,1.1) {};
        \node[font=\small] at (3.9,-0.3) {$g_t$};
        \node[sgn, draw=orange!80!black, fill=orange!25, very thick] (st) at (3.9,-0.95) {$\sigma_t$};
        \foreach \x/\i in {5.5/{t+1}, 7.3/n} {
            \node[col, draw=gray!60, dashed, fill=gray!5] at (\x,1.1) {};
            \node[font=\small, text=gray] at (\x,-0.3) {$g_{\i}$};
            \node[sgn, draw=gray!60, dashed] at (\x,-0.95) {};
        }
        \node[text=gray] at (6.4,1.1) {$\cdots$};
        \draw[->, thick, gray!70!black] (-0.3,3.15) -- (7.75,3.15) node[right, lab, text=black] {time};
        \node[lab, above] at (3.7,3.15) {the columns of $G$ arrive one at a time};
        \draw[decorate, decoration={brace, amplitude=4pt}] (-0.3,2.3) -- (3.0,2.3) node[midway, above=4pt, lab] {already seen};
        \node[lab, above, text=orange!70!black] at (3.9,2.3) {arriving now};
        \draw[decorate, decoration={brace, amplitude=4pt}, gray] (5.2,2.3) -- (7.6,2.3) node[midway, above=4pt, lab, text=gray!80!black] {not yet revealed};
        \draw[decorate, decoration={brace, amplitude=4pt, mirror}] (-0.3,-1.3) -- (3.0,-1.3) node[midway, below=4pt, lab] (fixed) {signs already fixed, never revised};
        \node[draw, rounded corners, thick, minimum width=1.2cm, minimum height=0.7cm] (A) at (3.9,-2.75) {$\subA_t$};
        \draw[->, thick] (fixed.south) |- (A.west);
        \draw[->, thick, orange!80!black] (cur.east) .. controls (5.0,0.4) and (5.0,-2.75) .. (A.east);
        \node[lab] (kw) at (3.9,-3.75) {$\margin$};
        \draw[->, thick] (kw) -- (A.south);
        \draw[->, thick] (A.north) -- (st.south);
    \end{tikzpicture}
    \caption{An online algorithm at step $t$. The columns of $G$ arrive one at a time, and each sign is chosen irrevocably when its column arrives. When $g_t$ arrives, the signs $\sigma_1, \dots, \sigma_{t-1}$ are already fixed, and $\subA_t$ chooses $\sigma_t$ using $g_t$, the earlier columns and signs, and the margin $\margin$. The later columns $g_{t+1}, \dots, g_n$ have not yet been revealed.}
    \label{fig:online}
\end{figure}

\begin{definition}[Online algorithms]\label{def:online}
    An online algorithm $\A$ takes as input a matrix $G$ with columns $g_1, \dots, g_n \in \R^m$ and a real number $\margin \in \R$. It outputs a string $\sigma \in \signs^n$, where for each $t \in [n]$, $\sigma_t$ is computed by the (perhaps randomized) measurable algorithm $\subA_t$ as
    \begin{align}
        \sigma_t \gets \subA_t(g_t; g_1 \| \cdots \| g_{t-1}, \sigma_1 \cdots \sigma_{t-1}, \margin).
    \end{align}
    Let $\online$ denote the class of all such algorithms. See \Cref{fig:online} for an illustration. Let $\online_K$ be the class of online algorithms defined as above, except that they receive a set $K \subseteq \R$ instead of $\margin$.
\end{definition}

\begin{definition}[Well-separated]
    A set $K \subseteq \R$ is \emph{well-separated} if it is a union of intervals (each of which may be open, closed, half-open, or a single point) such that any two intervals are at distance at least $\Delta$, for some $\Delta > 0$.
\end{definition}

\section{Main result and technical overview}\label{sec:technical-overview}

\begin{theorem}[Binary perceptron]\label{thm:main}
    Fix $\alpha \in (0,\infty)$, $\margin \in \R$, and let $m = \lfloor \alpha n \rfloor$. Let $h = \omega(\log^4 n)$. Then for any $\A \in \online$,
    \begin{align}
        \Prob{\A(G, \margin) \in \isosolset} < o(1),
    \end{align}
    where the probability is over $G \sim \G^{m,n}$ (a matrix with i.i.d.\ standard Gaussian entries) and the internal randomness of $\A$.
\end{theorem}

\begin{theorem}[General perceptrons]\label{thm:general-intro}
    Fix $\alpha \in (0,\infty)$, a well-separated set $K \subseteq \R$, and let $m = \lfloor \alpha n \rfloor$. Let $h = \omega(\log^4 n)$. Then for any $\A \in \online_K$,
    \begin{align}
        \Prob{\A(G, K) \in \mathcal{S}^h(G, K)} < o(1),
    \end{align}
    where the probability is over $G \sim \G^{m,n}$ and the internal randomness of $\A$.
\end{theorem}

Taking $K = [\margin, \infty)$ recovers \Cref{thm:main}, and taking $K = [-\margin, \margin]$ gives the $\mathsf{SBP}$. The intervals forming $K$ may be arbitrarily short, and there may be infinitely many of them, as in the square wave perceptron~\cite{BBMPRSZ25} where $K$ is periodic, i.e., $K = \bigcup_{k \in \mathbb{Z}} [2k, 2k+1]$. The theorem is only meaningful when isolated solutions exist. For example, if $K = \R$, then every $\sigma \in \signs^n$ is a solution, so there are no $h$-isolated solutions and the statement holds trivially.

\subsection{High-level proof strategy}

Our goal is to prove that every online algorithm successfully finds an isolated binary perceptron solution with probability at most $o(1)$. Recall that any online algorithm $\A$ uses a collection of algorithms $\subA_1, \dots, \subA_n$ to iteratively compute the output. An averaging argument (\Cref{lem:randomized}) means it suffices to assume that these algorithms are deterministic. Note, however, that each $\subA_t$ can compute a distinct function --- $\A$ can ``learn'' from its history to make decisions, so the choice of sign at step $t$ may be a function not only of the $t^{\text{th}}$ column, but also of all the prior $t - 1$ steps.

Our first key idea is to consider a seemingly sub-optimal algorithmic strategy: at a certain point, fix the last few sub-algorithms to compute \emph{the same function} $\parallelfunc: \R^m \to \signs$. Intuitively, such an algorithm essentially \emph{stops learning} while there are still several steps remaining. We show any online algorithm that successfully outputs an isolated solution must in fact behave this way.\footnote{
    Intuitively, this is because once there are fewer than $h$ columns remaining, there is at most one $h$-isolated solution that $\A$ can output, so there is nothing more for it to learn.
} Indeed, we show that for any successful online algorithm $\A$, there exists some integer $r$ and function $\parallelfunc$ such that (with high probability) the last $r$ signs of the output of $\A$ are computed by $\parallelfunc$. Since the output of a successful algorithm will be a perceptron solution, there will exist some $\residual \in \R^m$ (based on the threshold $\margin \sqrt{n}$) such that $\parallelfunc$ will pick out signs $\solsuff_i = \parallelfunc(\Gsuff_i)$ that satisfy
\begin{align}
    \score := \frac{1}{\sqrt{r}}\sum_{i=1}^r \solsuff_i \Gsuff_i \geq \residual. \label{line:score-inequality}
\end{align}

Our second key observation is that for small enough $r$, isolation forces $z$ to be the unique string in $\signs^r$ that satisfies \Cref{line:score-inequality}, which implies that $\score$ must lie in a narrow $m$-dimensional ``shell'' with high probability. A quantitative version of the central limit theorem implies that a Gaussian random variable $Z$ must also fall within the shell, which can only occur with probability $o(1)$ due to an anti-concentration property of Gaussian random variables. Therefore $\A$ can only succeed with probability $o(1)$.

\subsection{Successful algorithms must match a suffix-parallel baseline}\label{sec:success-close-to-baseline}

\begin{figure}[t]
    \centering
    \begin{tikzpicture}[
        lab/.style={font=\footnotesize},
        leaf/.style={circle, draw=orange!80!black, fill=orange!25, thick, inner sep=1.6pt}]
        \draw[line width=1.6pt, blue!60!black] (0,0) -- (6,0);
        \foreach \x in {0.5,1,...,5.5} \draw[blue!60!black, thick] (\x,-0.12) -- (\x,0.12);
        \node[lab, text=blue!60!black, above=4pt] at (3,0) {signs $\sigma'$ already fixed};
        \node[lab, align=center, below=4pt] at (3,0) {prefix $\Gpref$: the first $n-r$ columns,\\ signed one at a time};
        \node[circle, fill=black, inner sep=2pt] (o) at (6,0) {};
        \pgfmathsetmacro{\branchWidth}{3.2*cos(45)}
        \foreach \a/\i in {45/1, 22.5/2, 0/3, -45/r} {
            \draw[thick, orange!80!black] (o) -- ++(\branchWidth,{\branchWidth*tan(\a)}) node[leaf] (y\i) {}
                node[lab, pos=0.7, sloped, above] {$\Gsuff_{\i}$};
            \node[lab, right=2pt] at (y\i) {$\Gsuff_{\i}$};
        }
        \node at ($(o)+(-22.5:2.9)$) {$\vdots$};
    \end{tikzpicture}
    \caption{The suffix-parallel baseline $\A^{(r)}$ signs each suffix column $\Gsuff_i$ with the sign that $\A$ would choose were $\Gsuff_i$ to appear immediately after the prefix $\Gpref \in \R^{m \times (n-r)}$. \Cref{thm:successful-A-matches-baseline} determines an $r$ for which any successful $\A$ matches its baseline $\A^{(r)}$.}
    \label{fig:baseline}
\end{figure}
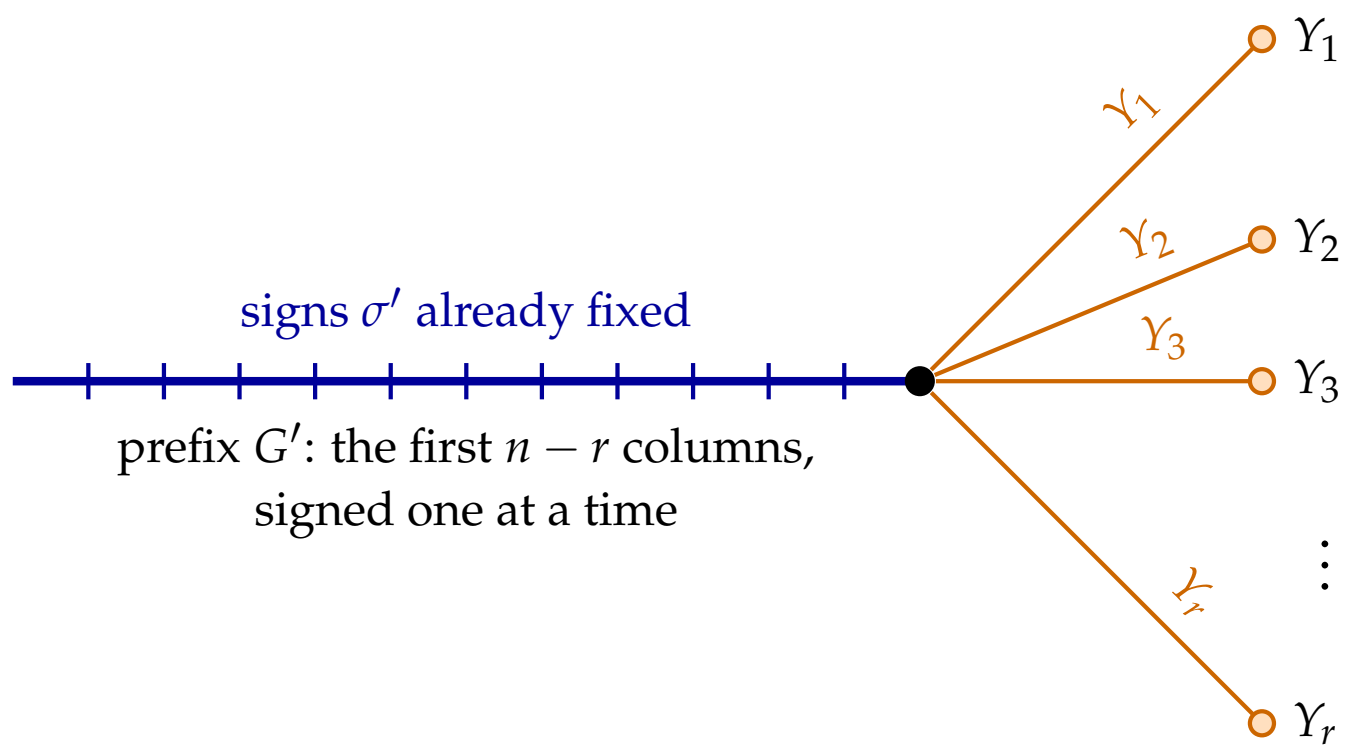

For any deterministic online algorithm $\A$, we consider the algorithm $\A^{(r)}$ that behaves identically to $\A$ up through step $n - r$. After this point, $\A^{(r)}$ will sign each of the last $r$ ``suffix'' columns as if they appeared in parallel at step $n - r + 1$ (see \Cref{fig:baseline}). 

\begin{definition}[Suffix-parallel baseline]\label{def:next-col-rule}
    Fix any deterministic $\A \in \online$ and integer $0 \leq r < n$. The \emph{$r$-suffix-parallel baseline for $\A$} is the algorithm $\A^{(r)} \in \online$ defined by
    \begin{align}
        \subA^{(r)}_t &:= \subA_t \qquad \quad  \forall t \in [n - r] \\
        \subA^{(r)}_t(\Gsuff; \cdot, \cdot, \cdot) &:= \subA_{n - r + 1}(\Gsuff; \Gpref, \sigma', \margin) \quad \forall t \in (n - r, n],
    \end{align}
    where $\Gpref \in \R^{m \times (n - r)}$ is the first $n - r$ columns of the input matrix to $\A$, $\sigma' \in \signs^{n-r}$ is the first $n-r$ signs of the solution committed to by $\A$ after processing $\Gpref$, and $\margin$ is the input margin to $\A$.
\end{definition}

Our argument holds that any successful online algorithm must ``stop learning'' and instead match some $r$-suffix-parallel baseline.

\begin{lemma}[Informal version of \Cref{thm:successful-A-matches-baseline}]\label{thm:informal-success-baseline}
    For any deterministic $\A \in \online$ and $\margin \in \R$, there exists some $r < h$ such that the following holds. With high probability over $G \sim \G^{m,n}$, if $\A$ successfully outputs an $h$-isolated solution to $G$, then $\A(G, \margin) = \A^{(r)}(G, \margin)$.
\end{lemma}

To prove \Cref{thm:successful-A-matches-baseline}, we consider any $r < h$ and sample $G \sim \G^{m,n}$. Let $\sigma' \in \signs^{n-r}$ be the solution prefix committed to by $\A$ after it processed the prefix columns $\Gpref \subset G$. Since $r < h$, there can be at most one $\solsuff \in \signs^r$ such that $\sigma' \| \solsuff$ is an isolated solution to $G$. Conditioning on the success of $\A$ implies that a unique such $z$ must exist, and for each suffix column $\Gsuff_i$ we have $\subA_{n-r+i}(\Gsuff_i) = z_i$.

It then suffices to show that the suffix-parallel baseline outputs $\sigma' \| z$. In particular, we need to show that $\subA^{(r)}_{n-r+i}(\Gsuff_i) = \subA_{n-r+1}(\Gsuff_i) = z_i$ for each $\Gsuff_i$.\footnote{
    Crucially: the subscript of $\subA$ is $n - r + 1$, not $n - r + i$.
} The crucial insight is that any successful $\A$ should choose the correct sign for $\Gsuff_i$ \emph{regardless of where $\Gsuff_i$ appears in the suffix}, since the distribution of $G$ is preserved when we exchange any two suffix columns. In particular, $\A$ should choose the correct sign when $\Gsuff_i$ appears immediately after $\Gpref$, which is exactly what is meant by $\subA_{n-r+1}(\Gsuff_i) = z_i$. We use this exchangeability observation to give an upper bound on the probability of the event $\{\A(G, \margin) \text{ is successful}\} \cap \{\A(G, \margin) \neq \A^{(r)}(G, \margin)\}$, for any $r < h$ within an asymptotically-growing interval. We then use an averaging argument to (implicitly) show the existence of some $r$ where this probability is $o(1)$.

\subsection{No suffix-parallel baseline can give a unique solution}\label{sec:baseline-not-feasible}

As we sketched above, for any input matrix $G$, when $r < h$, the suffix of the output of a successful $\A$ must be some $\solsuff \in \signs^r$ that \emph{uniquely} completes the solution prefix $\sigma'$ to create an isolated solution of $G$. Moreover, \Cref{thm:successful-A-matches-baseline} implies that, except on an event of probability $o(1)$, there exists a function $\nextcol := \subA_{n-r+1}$ such that $\nextcol(\Gsuff_i) = \solsuff_i$ for all $i \in [r]$. In other words, $\nextcol$ produces the unique string $\solsuff \in \signs^r$ such that
\begin{align}
    \frac{1}{\sqrt{r}} \sum_{i=1}^r \solsuff_i \Gsuff_i \geq \frac{\margin \sqrt{n} \mathbf{1}_m - \Gpref \sigma'}{\sqrt{r}}. \label{line:z-satisfies}
\end{align}

Note that $\nextcol$ is (implicitly) a function of $\A$ and the prefix columns $\Gpref$, but is independent of the suffix columns $\Gsuff_1, \dots, \Gsuff_r$. In \Cref{pf:sec:baseline-not-feasible} we prove that no function $\parallelfunc$ can produce a $z \in \signs^r$ that uniquely satisfies \Cref{line:z-satisfies}, except with probability $o(1)$ over $\Gsuff_1, \dots, \Gsuff_r$.

\begin{definition}[Uniqueness event]\label{def:unique-event}
    For any $m, r \in \N$, let $\parallelfunc: \R^m \to \signs$ be any measurable function and sample $\Gsuff_1, \dots, \Gsuff_r \sim \stdnormal$ independently. For any $\residual \in \R^m$, define $\unique$ as the event that the vector $z \in \signs^r$ given by $z_i = \parallelfunc(\Gsuff_i)$ is the unique element of $\signs^r$ such that
    \begin{align}
        \frac{1}{\sqrt{r}}\sum_{i=1}^r z_i\Gsuff_i \geq \residual.
    \end{align}
\end{definition}

\Cref{thm:anti-uniqueness} states that for any $\parallelfunc: \R^m \to \signs$ and any $\residual \in \R^m$, we have
\begin{align}
    \Prob{\unique} \leq O\left(\frac1m + \frac{(\log m)^{3/2} \log r + (\log m)^2}{\sqrt{r}} + \frac{\log m}{\sqrt{r}}\right), \label{line:informal-prob-anti-unique}
\end{align}
with the probability taken over $\Gsuff_1, \dots, \Gsuff_r \sim \stdnormal$ independently. We will ultimately apply \mbox{\Cref{thm:anti-uniqueness}} with $m > \omega(1)$ and $r > \omega(\log^4 m)$, which will imply that the probability in \mbox{\Cref{line:informal-prob-anti-unique}} is $o(1)$.

For any function $\parallelfunc$, we denote its score $\score \in \R^m$ on the suffix columns $\Gsuff_1, \dots, \Gsuff_r$ by
\begin{align}
    \score := \frac{1}{\sqrt{r}}\sum_{i=1}^r \parallelfunc(\Gsuff_i)\Gsuff_i.
\end{align}
The uniqueness constraint of $\unique$ implies that with high probability, $\score$ will lie in a narrow ``shell'' in $\R^m$ (see \Cref{fig:shell-gaussian-shell} or \Cref{lem:unique-to-orthant-switch}). Since $\score$ is a (scaled) sum of independent, identically distributed random variables, we can use the Berry-Esseen theorem (\Cref{thm:berry-esseen}) to instead analyze a Gaussian random variable $Z$ with the same variance as $\score$. We then finish the proof by appealing to an anti-concentration property of Gaussian random variables (see \Cref{fig:shell-gaussian-normal} or \Cref{thm:nazarov}), showing that $Z$ will fall into such a narrow shell with probability at most $o(1)$. For the last two steps, the key technical piece is to show that the variance of $\score$ is bounded above and below by positive constants, which follows because $\parallelfunc$ is bounded and independent of $\Gsuff_1, \dots, \Gsuff_r$ (\Cref{lem:hypotheses-for-berry-esseen}).

In more detail, the proof of \Cref{thm:anti-uniqueness} is composed of three conceptual moves, each associated with one of the three terms in the upper bound of \Cref{line:informal-prob-anti-unique}. Our first move is to show that $\unique$ implies that $\score$ lies within a small shell of width $\delta = \Theta(\sqrt{(\log m) / r})$, except on an event $\mathcal{E}_m^c$ that occurs with probability $O(1/m)$. In particular, $\score$ will lie in the difference of two upper orthants in $\R^m$, i.e.,
\begin{align}
    \unique \cap \mathcal{E}_m \subseteq \{\score \in \orthant{\residual} \setminus \orthant{\residual + \delta\mathbf{1}_m}\}. \label{line:subset-line}
\end{align}
Recall that $\orthant{\residual}$ is the set of $x \in \R^m$ such that $x_j \geq \residual_j$ for all $j \in [m]$. The event $\unique$ implies that $\score \geq \residual$, so $\score \in \orthant{\residual}$. To show additionally that $\score \notin \orthant{\residual + \delta\mathbf{1}_m}$ requires finding some $j^*$ such that $(\score)_{j*} < \residual_{j*} + \delta$. The uniqueness constraint of $\unique$ implies that flipping the sign of $\parallelfunc(\Gsuff_1)$ would force $(\score)_{j^*} < \residual_{j^*}$ for some $j^* \in [m]$. We use the union bound and a standard Gaussian tail bound to show that the event $\mathcal{E}_m := \{|\Gsuff_{1j}| \leq O(\sqrt{\log m}) \text{ for all } j \in [m]\}$ occurs with high probability. Therefore flipping the sign of $\parallelfunc(\Gsuff_1)$ cannot change the value of $\score$ by more than $O(\sqrt{(\log m)/r})$ at each row. Therefore it must be the case that $(\score)_{j^*} < \residual_{j^*} + \delta$.

\begingroup

\def\thetaOne{2.5}
\def\thetaTwo{1.2}
\def\deltaLength{1.5}
\def\deltaLengthGaussian{1.5}
\def\xMax{8}
\def\yMax{4}
\colorlet{thetaColor}{black}
\colorlet{deltaColor}{black}
\colorlet{shadeColor}{blue!20}
\colorlet{symbolColor}{blue}
\newif\ifshowS
\showSfalse %

\def\orthantFill{none}        %
\def\shiftedOrthantFill{none} %
\def\shellOpacity{1}         %

\newif\ifshowPoints
\showPointstrue              %
\newif\ifhorizontalConnection
\horizontalConnectionfalse  %
\def\pointX{5.2}             %
\def\pointY{2}
\def\outsideGap{.5}         %
\def\pointRadius{3pt}
\def\insideLabel{$S_\phi$}
\def\outsideLabel{$S'_\phi$}
\def\insideLabelAnchor{west}
\def\outsideLabelAnchor{west}
\def\insideLabelXShift{3pt}  %
\def\insideLabelYShift{0pt}
\def\outsideLabelXShift{3pt}
\def\outsideLabelYShift{0pt}
\colorlet{insideDotColor}{black}
\colorlet{outsideDotColor}{black}
\colorlet{insideLabelColor}{black}
\colorlet{outsideLabelColor}{black}
\colorlet{connectionColor}{black}

\def\gaussianMean{0}
\def\gaussianSD{1.4}
\def\gaussianStart{0}

\newif\ifshowBound
\showBoundtrue              %
\def\boundText{$\leq \frac{\delta}{\sigma}\left(\sqrt{2\log m}+2\right)$}
\def\boundLabelX{1.1}
\def\boundLabelY{.35}
\def\boundTargetX{.85}       %
\def\boundTargetY{.16}
\def\boundArrowOut{225}     %
\def\boundArrowIn{75}
\colorlet{boundLabelColor}{black}
\colorlet{boundArrowColor}{black}

\def\shellCaption{The shell $\orthant{\residual} \setminus \orthant{\residual + \delta\mathbf{1}_m}$ in $m=2$ dimensions. \Cref{lem:unique-to-orthant-switch} shows that $\score$ will lie in this shell, since changing any sign to give $\score'$ must drop it below $\orthant{\residual}$.}

\def\gaussianCaption{Nazarov's inequality in $m=1$ dimension. Intuitively, the upper bound should grow with $\delta$ and decrease with the variance $\sigma^2$.}

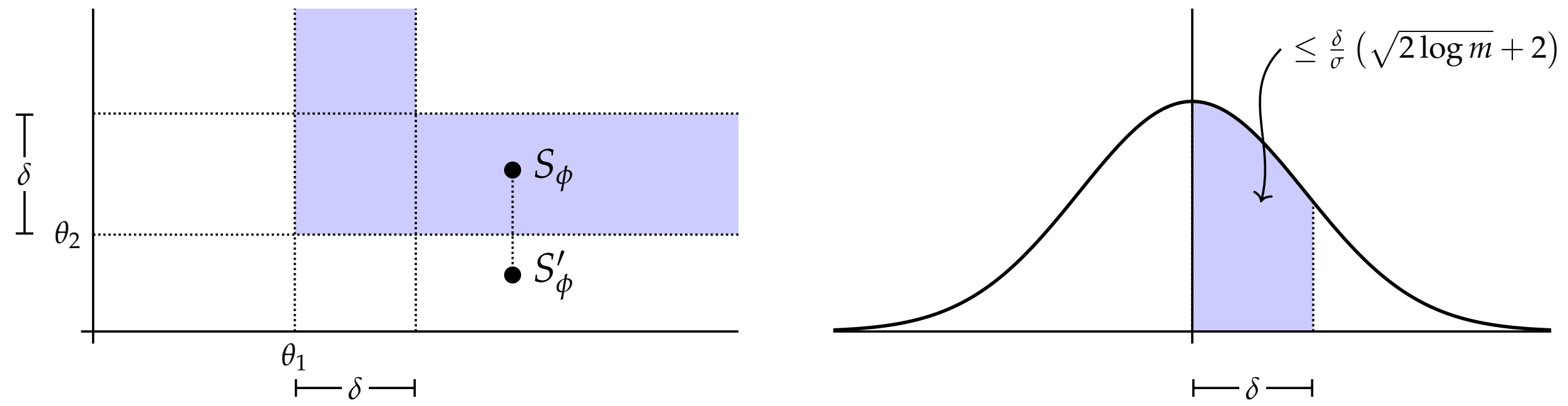
\def\overallCaption{Helpful figures for the proof of \Cref{thm:anti-uniqueness}. The Berry-Esseen theorem states that the probability of $\score$ lying in the width-$\delta$ shell is close to the probability that a Gaussian $Z$ with the same covariance lies in the shell. Nazarov's inequality says that the probability that any such $Z$ lies in the shell is at most $\frac{\delta}{\sigma}\left(\sqrt{2\log m} + 2\right)$, where $\sigma^2 > 0$ is a lower bound on each $\Var(Z_j)$.}

\pgfmathsetmacro{\panelHalfWidth}{(\xMax+1.3)/2}
\pgfmathsetmacro{\gaussianRange}{\panelHalfWidth-.2}
\pgfmathsetmacro{\gaussianEnd}{\gaussianStart+\deltaLengthGaussian}

\begin{figure}[ht]
\centering
\begin{subfigure}[t]{.48\linewidth}
\vspace{0pt}
\centering
\resizebox{\linewidth}{!}{%
\begin{tikzpicture}[
    x=1cm, y=1cm, thick,
    guide/.style={densely dotted}
]
    \path[fill=\orthantFill]
        (\thetaOne,\thetaTwo) rectangle (\xMax,\yMax);
    \path[fill=\shiftedOrthantFill]
        ({\thetaOne+\deltaLength},{\thetaTwo+\deltaLength})
        rectangle (\xMax,\yMax);

    \fill[shadeColor,fill opacity=\shellOpacity]
        (\thetaOne,\thetaTwo)
        -- (\xMax,\thetaTwo)
        -- (\xMax,{\thetaTwo+\deltaLength})
        -- ({\thetaOne+\deltaLength},{\thetaTwo+\deltaLength})
        -- ({\thetaOne+\deltaLength},\yMax)
        -- (\thetaOne,\yMax) -- cycle;

    \draw[guide,thetaColor]
        (\thetaOne,0) -- (\thetaOne,\yMax)
        (0,\thetaTwo) -- (\xMax,\thetaTwo);
    \draw[guide,deltaColor]
        ({\thetaOne+\deltaLength},0)
            -- ({\thetaOne+\deltaLength},\yMax)
        (0,{\thetaTwo+\deltaLength})
            -- (\xMax,{\thetaTwo+\deltaLength});

    \draw (-0.15,0) -- (\xMax,0);
    \draw (0,-0.15) -- (0,\yMax);
    \node[below] at (\thetaOne,0) {$\residual_1$};
    \node[left]  at (0,\thetaTwo) {$\residual_2$};

    \draw[|-|] (\thetaOne,-0.7)
        -- node[fill=white,inner sep=2pt] {$\delta$}
        ({\thetaOne+\deltaLength},-0.7);
    \draw[|-|] (-0.85,\thetaTwo)
        -- node[fill=white,inner sep=2pt] {$\delta$}
        (-0.85,{\thetaTwo+\deltaLength});

    \ifshowS
        \node[text=symbolColor,font=\Large]
            at ({\thetaOne+\deltaLength/2},{\thetaTwo+\deltaLength/2})
            {$S_\phi$};
    \fi

    \ifshowPoints
        \coordinate (inside) at (\pointX,\pointY);
        \ifhorizontalConnection
            \coordinate (outside) at ({\thetaOne-\outsideGap},\pointY);
        \else
            \coordinate (outside) at (\pointX,{\thetaTwo-\outsideGap});
        \fi
        \draw[guide,connectionColor] (inside) -- (outside);
        \fill[insideDotColor]  (inside)  circle[radius=\pointRadius];
        \fill[outsideDotColor] (outside) circle[radius=\pointRadius];
        \node[anchor=\insideLabelAnchor,text=insideLabelColor,font=\Large,
              xshift=\insideLabelXShift,yshift=\insideLabelYShift]
            at (inside) {\insideLabel};
        \node[anchor=\outsideLabelAnchor,text=outsideLabelColor,font=\Large,
              xshift=\outsideLabelXShift,yshift=\outsideLabelYShift]
            at (outside) {\outsideLabel};
    \fi

    \pgfresetboundingbox
    \path[use as bounding box]
        (-1.1,-1.05) rectangle ({\xMax+.2},{\yMax+.2});
\end{tikzpicture}%
}
\caption{\shellCaption}
\label{fig:shell-gaussian-shell}
\end{subfigure}\hfill
\begin{subfigure}[t]{.48\linewidth}
\vspace{0pt}
\centering
\resizebox{\linewidth}{!}{%
\begin{tikzpicture}[
    x=1cm, y=10cm, thick,
    guide/.style={densely dotted},
    declare function={
        normalpdf(\x)=exp(-(\x-\gaussianMean)^2/(2*\gaussianSD^2))
                      /(\gaussianSD*sqrt(2*pi));
    }
]
    \fill[shadeColor]
        (\gaussianStart,0)
        -- plot[domain=\gaussianStart:\gaussianEnd,samples=101]
            (\x,{normalpdf(\x)})
        -- (\gaussianEnd,0) -- cycle;
    \draw[guide,thetaColor]
        (\gaussianStart,0) -- (\gaussianStart,{normalpdf(\gaussianStart)});
    \draw[guide,deltaColor]
        (\gaussianEnd,0) -- (\gaussianEnd,{normalpdf(\gaussianEnd)});
    \draw[very thick]
        plot[domain=-\gaussianRange:\gaussianRange,samples=151,smooth]
            (\x,{normalpdf(\x)});

    \draw (-\gaussianRange,0) -- (\gaussianRange,0);
    \draw (0,-.015) -- (0,{\yMax/10});
    \draw[|-|] (\gaussianStart,-.07)
        -- node[fill=white,inner sep=2pt] {$\delta$}
        (\gaussianEnd,-.07);

    \ifshowBound
        \node[anchor=west,text=boundLabelColor] (boundLabel)
            at (\boundLabelX,\boundLabelY) {\boundText};
        \draw[->,boundArrowColor]
            (boundLabel.west) to[out=\boundArrowOut,in=\boundArrowIn]
            (\boundTargetX,\boundTargetY);
    \fi

    \pgfresetboundingbox
    \path[use as bounding box]
        (-\panelHalfWidth,-.105)
        rectangle (\panelHalfWidth,{(\yMax+.2)/10});
\end{tikzpicture}%
}
\caption{\gaussianCaption}
\label{fig:shell-gaussian-normal}
\end{subfigure}
\caption{\overallCaption}
\label{fig:shell-gaussian}
\end{figure}
\endgroup

Our second move is to punt on analyzing the probability of $\{\score \in \orthant{\residual} \setminus \orthant{\residual + \delta\mathbf{1}_m}\}$ and instead sample the multidimensional Gaussian $Z \sim \mathcal{N}(\sqrt{r}\mu, \Sigma)$ and estimate the probability of $\{Z \in \orthant{\residual} \setminus \orthant{\residual + \delta\mathbf{1}_m}\}$.\footnote{
    Here, $\mu := \Ex{\parallelfunc(\Gsuff_1)\Gsuff_1}$ and $\Sigma := \Cov(\score)$, so $Z$ has the same mean and covariance as $\score$. In fact, in our proof we analyze the centered version of $Z$, but present the non-centered version here for clarity.
} To justify this, we use a version of the Berry-Esseen theorem (\Cref{thm:berry-esseen}), which one can think of as a quantitative version of the central limit theorem. In particular, we show that
\begin{align}
    \left|\Prob{\score \in \orthant{\residual} \setminus \orthant{\residual + \delta\mathbf{1}_m}} - \Prob{Z \in \orthant{\residual} \setminus \orthant{\residual + \delta\mathbf{1}_m}}\right| \leq O\left(\frac{(\log m)^{3/2}\log r + (\log m)^2}{\sqrt{r}}\right), \label{line:berry-esseen-line}
\end{align}
relying crucially on the fact that $\Var(Z_j) = 1 - \mu_j^2$ is bounded above and below by a positive constant.\footnote{
    This is not obvious, but is shown in \Cref{lem:hypotheses-for-berry-esseen}.
}

Finally, we use Nazarov's inequality (\Cref{thm:nazarov}) to show an anti-concentration property of $Z$. In particular,
\begin{align}
    \Prob{Z \in \orthant{\residual} \setminus \orthant{\residual + \delta\mathbf{1}_m}} \leq O\left(\frac{\delta}{\sigma}\sqrt{\log m}\right) = O\left(\frac{\log m}{\sqrt{r}}\right) \label{line:nazarov-line}
\end{align}
which again relies on the fact that $\sigma^2 := \min_j\{\Var(Z_j)\}$ is bounded below by some positive constant. Combining \Cref{line:subset-line,line:berry-esseen-line,line:nazarov-line} suffices to prove \Cref{thm:anti-uniqueness}. 

\subsection{Putting it all together}\label{sec:all-together}

\begin{theorem}\label{thm:main-deterministic}
    Fix $\alpha \in (0, \infty)$ and let $m = \lfloor \alpha n \rfloor$. Fix any deterministic $\A \in \online$ and any $\margin \in \R$. Then for any function $h(n)$ such that $\omega(\log^4 n) < h(n) < n$, we have
    \begin{align}
        \Prob{\A(G, \margin) \in \isosolset} < o(1),
    \end{align}
    with the probability taken over $G \sim \G^{m,n}$. Moreover, this bound does not depend on $\A$.
\end{theorem}
\begin{proof}
    Fix an integer $r$ to be specified later, and define the algorithm $\A^{(r)}$. Sample $G \sim \G^{m,n}$, let $t(n) = h(n)/\log^4 n$, and define $a(n) := \lceil \sqrt{t(n)}\log^4 n\rceil $. 
    Define the events
    \begin{align}
        \success &:= \{\A(G, \margin) \in \isosolset\} \\
        \match_r &:= \{\A(G, \margin) = \A^{(r)}(G, \margin)\}.
    \end{align}
    \Cref{thm:successful-A-matches-baseline} states that there exists some $r$ satisfying $a \leq r < h$ such that
    \begin{align}
        \Prob{\success} &= \Prob{\success \cap \match_r} + \Prob{\success \cap \match_r^c} \\
        &< \Prob{\success \cap \match_r} + o(1).
    \end{align}
    We now analyze the event $\success \cap \match_r$. Denote by $\Gpref$ the first $n - r$ columns of $G$ and denote the last $r$ columns by $\Gsuff_1, \dots, \Gsuff_r$. Write $\A(G, \margin) = \sigma' \| \solsuff$, for some $z \in \signs^r$. Define
    \begin{align}
        \residual := \frac{\margin \sqrt{n}\mathbf{1}_m - G'\sigma'}{\sqrt{r}}.
    \end{align}
    We claim that $\success \cap \match_r \subseteq \mathcal{U}_{\nextcol, \residual}$. To see why, note that $\success$ and the fact that $r < h$ together imply that $\solsuff$ is the unique string in $\signs^r$ that satisfies $r^{-1/2}\sum_{i=1}^r \solsuff_i \Gsuff_i \geq \residual$. Let $\nextcol := \subA_{n-r+1}$, so $\match_r$ implies that $\nextcol(\Gsuff_i) = \solsuff_i$ for each $i \in [r]$.\footnote{
        By definition, the last $r$ signs of the output of $\A^{(r)}$ are given by $\nextcol(\Gsuff_i)$.
    } Conditional on $\Gpref$, the columns $\Gsuff_1, \dots, \Gsuff_r$ are independent with distribution $\stdnormal$, while $\nextcol$ and $\residual$ are fixed. Therefore we have
    \begin{align}
        \Prob{\success \cap \match_r} &\leq \Prob{\mathcal{U}_{\nextcol,\residual}} \\
        &\leq O\left(\frac1m + \frac{(\log m)^{3/2} \log r + (\log m)^2}{\sqrt{r}} + \frac{\log m}{\sqrt{r}}\right) \tag{\Cref{thm:anti-uniqueness}} \\
        &< o(1),
    \end{align}
    with the last inequality due to $m > \omega(1)$ and $r \geq a(n) > \omega(\log^4 m)$. More precisely, since $x \mapsto (\log x)/\sqrt{x}$ is decreasing for $x \geq e^2$, the bound is largest at $r = a(n)$, where each term tends to zero because $\sqrt{a(n)} \geq t(n)^{1/4}\log^2 n$ and $\log a(n) = O(\log t(n) + \log\log n)$. These bounds hold for every $r \geq a(n)$ and otherwise do not depend on $\A$. The result follows.
\end{proof}

\Cref{thm:main} follows from \Cref{thm:main-deterministic}. Indeed, since the bound in \Cref{thm:main-deterministic} does not depend on $\A$, \Cref{lem:randomized} extends it to all algorithms in $\online$. Moreover, if $h(n) \geq n$, we may replace $h$ by $n - 1$, since every $h$-isolated solution is also $h'$-isolated for $h' \leq h$.

\section{Main proofs}\label{sec:proofs}

\subsection{Successful algorithms must match a suffix-parallel baseline}\label{pf:sec:success-close-to-baseline}

\begin{lemma}\label{thm:successful-A-matches-baseline}
    Fix any deterministic $\A \in \online$ and $\margin \in \R$. For any $0 \leq r < n$, define the $r$-suffix-parallel baseline $\A^{(r)}$. Sample $G \sim \G^{m,n}$ and write $G = \Gpref \| \Gsuff_1 \| \cdots \| \Gsuff_r$, where $\Gpref \in \R^{m \times (n-r)}$ and $\Gsuff_i \in \R^m$ for each $i \in [r]$. For any $h = h(n)$ such that $\omega(1) < h < n$, define the events
    \begin{align}
        \success &:= \{\A(G, \margin) \in \isosolset\} \\
        \match_r &:= \{\A(G, \margin) = \A^{(r)}(G, \margin)\}.
    \end{align}
    Then for any integer-valued function $a = a(n)$ satisfying $1 \leq a < o(h)$, there exists an integer-valued function $r = r(n)$ satisfying $a \leq r < h$ such that
    \begin{align}
        \underProb{G}{\success \cap \match_r^c} \leq H^{-1}, \qquad \text{where } H := \sum_{r'=a}^{\lfloor h/2 \rfloor} \frac{1}{r'} = \Theta\left(\log (h/a)\right).
    \end{align}
    In particular, $\Prob{\success \cap \match_r^c} = o(1)$, and this bound does not depend on $\A$.
\end{lemma}

\begin{proof}
    Sample $G \sim \G^{m,n}$ and let $\sigma_1 \cdots \sigma_n \in \signs^n$ denote the output of $\A(G, \margin)$. For any $r \in \{0, 1, \dots, n\}$, define the event
    \begin{align}
        E_r := \{\exists \tau \in \isosolset: \tau_{1:n-r} = \sigma_{1:n-r}\}.
    \end{align}
    In other words, $E_r$ is the event that there exists some isolated solution whose prefix agrees with the partial solution $\sigma' = \sigma_{1:n-r}$ committed to by $\A$ when there are exactly $r$ steps remaining. Observe that $\success = E_0$ and for each $r$ we have $E_r \subseteq E_{r+1}$. Moreover, when $E_r$ occurs and $r < h$, there must be exactly one $\solsuff^* \in \signs^r$ such that $\sigma'\|\solsuff^*$ is an isolated solution, as otherwise $h$-isolation would be contradicted.

    Fix $r$ such that $1 \leq r < h$. Let $\Gsuff_1, \dots, \Gsuff_r$ denote the remaining suffix columns of $G$, and let $\nextcol := \subA_{n-r+1}$, so $\nextcol(\Gsuff_i)$ denotes the sign that $\A$  would choose if $\Gsuff_i$ would appear immediately after $\Gpref$. Since the map $\subA_{n-r+1}$ in \Cref{def:online} is measurable, so is $\parallelfunc_r$. Recall that for each $i \in [r]$, $\sigma_{n-r+i}$ denotes the actual sign chosen by $\A$ upon receipt of $\Gsuff_i$. Then by definition we have
    \begin{align}
        \match_r = \{\parallelfunc_r(\Gsuff_i) = \sigma_{n-r+i} \text{ for all } i \in [r]\}.
    \end{align}
    Recalling that $\solsuff^*$ is defined on $E_r$, define the event
    \begin{align}
        \mistake_{r,i} := E_r \cap \{\parallelfunc_r(\Gsuff_i) \neq \solsuff^*_i\},
    \end{align}
    i.e., that if $\A$ were to see the column $\Gsuff_i$ immediately after $G'$, it would fail to correctly sign $\Gsuff_i$. We claim that
    \begin{align}
        \success \cap \match_r^c \subseteq \bigcup_{i=1}^r \mistake_{r,i}. \label{line:success-no-match}
    \end{align}
    To see why, note that $\success$ implies $E_r$ and that $\sigma_{n-r+i} = \solsuff^*_i$ for all $i \in [r]$. Moreover, $\match_r^c$ implies that there must be some $i^* \in [r]$ for which $\parallelfunc_r(\Gsuff_{i^*}) \neq \sigma_{n - r + {i^*}} = \solsuff^*_{i^*}$, so $\mistake_{r,i^*}$ must occur. We now proceed to bound the probability of the right-hand side of \Cref{line:success-no-match}. First, we note that since the $\Gsuff_i$ are independent and identically distributed, exchanging the columns $\Gsuff_i$ and $\Gsuff_1$ merely swaps the values of $\solsuff^*_i$ and $\solsuff^*_1$, but does not affect $\Gpref$, nor the distribution of $G$, nor the event $E_r$. Therefore for all $i \in [r]$ we have
    \begin{align}
        \Prob{\mistake_{r,i}} = \Prob{\mistake_{r,1}}.
    \end{align}
    Moreover, we have
    \begin{align}
        \mistake_{r,1} = E_r \setminus E_{r-1} \label{line:mistake-means}
    \end{align}
    since if $\A$ chooses the wrong sign upon receipt of $\Gsuff_1$, the uniqueness of $\solsuff^*$ implies that $E_{r-1}$ cannot occur. In light of \Cref{line:success-no-match,line:mistake-means}, we can use a union bound to obtain
    \begin{align}
        \Prob{\success \cap \match_r^c} &\leq \sum_{i=1}^r \Prob{\mistake_{r,i}} \\
        &\leq r\Prob{\mistake_{r,1}} \\
        &= r(\Prob{E_r} - \Prob{E_{r-1}}). \label{line:upper-bound-on-no-match}
    \end{align}
    Choose any integer-valued function $a = a(n)$ such that $1 \leq a < o(h)$ and choose $\ell = \lfloor h/2\rfloor$, so for large enough $n$ we have $a < \ell$. Then from \Cref{line:upper-bound-on-no-match} we obtain
    \begin{align}
        \sum_{r = a}^\ell \frac{\Prob{\success \cap \match_r^c}}{r} &\leq \sum_{r = a}^\ell \left(\Prob{E_r} - \Prob{E_{r-1}}\right) \\
        &= \Prob{E_\ell} - \Prob{E_{a-1}} \\
        &\leq 1. \label{line:weighted-avg}
    \end{align}
    Define
    \begin{align}
        H := \sum_{r = a}^\ell \frac1r.
    \end{align}
    An averaging argument shows there exists some $r \in \{a, \dots, \ell\}$ such that $\Prob{\success \cap \match_{r}^c} \leq H^{-1}$, as otherwise we would have
    \begin{align}
        \sum_{r' = a}^\ell \frac{\Prob{\success \cap \match_{r'}^c}}{r'} > H^{-1} \sum_{r'=a}^\ell \frac{1}{r'} = 1,
    \end{align}
    which contradicts \Cref{line:weighted-avg}. Since $H$ is a partial harmonic series we have $H = \Theta(\log (\ell / a)) > \omega(1)$, and the result follows.
\end{proof}

\subsection{No suffix-parallel baseline can give a unique solution}\label{pf:sec:baseline-not-feasible}

In this section we prove \Cref{thm:anti-uniqueness}. We rely on \Cref{lem:unique-to-orthant-switch,lem:hypotheses-for-berry-esseen}, which are proved in Appendix~\ref{ap:lemmas-for-unique}.

\begin{lemma}[Upper bound on uniqueness]\label{thm:anti-uniqueness}
    Fix any integers $m, r \geq 3$, any measurable function $\parallelfunc: \R^m \to \signs$, and any $\residual \in \R^m$. Sample $\Gsuff_1, \dots, \Gsuff_r \sim \stdnormal$ independently and define $\unique$ as in \Cref{def:unique-event}. Then
    \begin{align}
        \Prob{\unique} \leq O\left(\frac1m + \frac{(\log m)^{3/2} \log r + (\log m)^2}{\sqrt{r}} + \frac{\log m}{\sqrt{r}}\right), \label{line:uniqueness-upper-bound}
    \end{align}
    with the probability taken over $\Gsuff_1, \dots, \Gsuff_r$. The implicit constant is absolute, so the bound is uniform over $\parallelfunc$ and $\residual$.
\end{lemma}

\begin{proof}
    For clarity, define $S_i := \parallelfunc(\Gsuff_i) \Gsuff_i$ for each $i \in [r]$ and let
    \begin{align}
        S := \frac{1}{\sqrt{r}} \sum_{i = 1}^r S_i.
    \end{align}
    By \Cref{lem:unique-to-orthant-switch}, there exists a $\delta = \Theta(\sqrt{(\log m)/r})$ such that
    \begin{align}
        \Prob{\unique} \leq \Prob{S \in \orthant{\residual} \setminus \orthant{\residual + \delta \mathbf{1}_m}} + O\left(\frac{1}{m}\right). \label{line:high-prob-event}
    \end{align}
    We now proceed to bound $\Prob{S \in \orthant{\residual} \setminus \orthant{\residual + \delta \mathbf{1}_m}}$. For each $i \in [r]$, write $\mu := \Ex{S_i}$ and define the centered random variable $W_i := S_i - \mu$, along with
    \begin{align}
        W := \frac{1}{\sqrt{r}} \sum_{i=1}^r W_i.
    \end{align}
    Let $\Sigma$ be the covariance matrix of $W$ and sample $Z \sim \mathcal{N}(0, \Sigma)$. Let 
    \begin{align}
        \rectangleset := \{\Pi_{j=1}^m [a_j, b_j): -\infty \leq a_j \leq b_j \leq \infty \text{ for every } j \in [m]\}
    \end{align}
    be the set of rectangles in $\R^m$ and note that $\orthant{u} \in \rectangleset$ for all $u \in \R^m$. By \Cref{lem:hypotheses-for-berry-esseen}, the hypotheses of \Cref{thm:berry-esseen} hold, which implies that
    \begin{align}
        \rho_{m, r} := \sup_{A \in \rectangleset} \left|\Prob{W \in A} - \Prob{Z \in A}\right| \leq O\left(\frac{(\log m)^{3/2}\log r + (\log m)^2}{\sqrt{r}}\right). \label{line:berry-esseen}
    \end{align}
    So for any $u \in \R^m$, the triangle inequality and \Cref{line:berry-esseen} give
    \begin{align}
        \left|\Prob{W \in \orthant{u} \setminus \orthant{u + \delta\mathbf{1}_m}} - \Prob{Z \in \orthant{u} \setminus \orthant{u + \delta\mathbf{1}_m}}\right| &= |(\Prob{W \in \orthant{u}} - \Prob{W \in \orthant{u + \delta\mathbf{1}_m}}) \\
        &- (\Prob{Z \in \orthant{u}} - \Prob{Z \in \orthant{u + \delta\mathbf{1}_m}})| \\
        &= |(\Prob{W \in \orthant{u}} - \Prob{Z \in \orthant{u}}) \\
        &- (\Prob{W \in \orthant{u + \delta\mathbf{1}_m}} - \Prob{Z \in \orthant{u + \delta\mathbf{1}_m}})| \\
        &\leq 2\rho_{m,r}.
    \end{align}
    Since $W = S - \sqrt{r}\mu$, we have
    \begin{align}
        \Prob{S \in \orthant{\residual} \setminus \orthant{\residual + \delta\mathbf{1}_m}} &= \Prob{W \in \orthant{\theta - \sqrt{r}\mu} \setminus \orthant{\residual - \sqrt{r}\mu + \delta\mathbf{1}_m}} \\
        &\leq \Prob{Z \in \orthant{\theta - \sqrt{r}\mu} \setminus \orthant{\residual - \sqrt{r}\mu + \delta\mathbf{1}_m}} + 2\rho_{m, r} \\
        &\leq \frac{\delta}{\sqrt{\varW}}\left(\sqrt{2 \log m} + 2\right) + 2\rho_{m,r}. \label{line:S-in-orthant}
    \end{align}
    The last inequality holds by an application of Nazarov's inequality (see \Cref{thm:nazarov}, which we can apply because $\Sigma_{jj} \geq \varW > 0$ for each $j \in [m]$, which is shown in \Cref{lem:hypotheses-for-berry-esseen}). Putting together \Cref{line:high-prob-event,line:berry-esseen,line:S-in-orthant}, we have
    \begin{align}
        \Prob{\unique} \leq O\left(\frac{1}{m}\right) + O\left(\frac{(\log m)^{3/2}\log r + (\log m)^2}{\sqrt{r}}\right) + \frac{\delta}{\sqrt{\varW}}\left(\sqrt{2\log m} + 2\right).
    \end{align}
    Recalling that $\delta = \Theta(\sqrt{(\log m)/r})$ and $\varW$ is a positive constant completes the proof.
\end{proof}

\subsection{General perceptrons}\label{sec:general-perceptrons}

In this section we prove \Cref{thm:general-intro}. The proof will follow that of \Cref{thm:main-deterministic}, which uses \Cref{thm:successful-A-matches-baseline,thm:anti-uniqueness}. We can use \Cref{thm:successful-A-matches-baseline} as-is. Indeed, its proof uses only two facts about isolated solutions, namely that an $h$-isolated solution is the only solution within Hamming distance $h$ of itself, and that exchanging suffix columns of $G$ together with the corresponding signs preserves the set of isolated solutions. For any set $K \subseteq \R$, both facts hold for $\mathcal{S}^h(G, K)$, so \Cref{thm:successful-A-matches-baseline} holds verbatim for deterministic algorithms in $\online_K$, with $K$ in place of $\margin$ and $\mathcal{S}^h(G, K)$ in place of $\isosolset$. The proof of \Cref{lem:randomized} also applies verbatim to $\online_K$.

However, we need to extend \Cref{thm:anti-uniqueness} from a statement about upper orthants --- which aligns with the binary perceptron threshold of $\margin \sqrt{n}$ --- to a statement about rectangles, to capture the general perceptron goal of coordinatewise inclusion in a set $K \subseteq \R$.

\begin{definition}[Rectangles]\label{def:rectangles}
    Given any set of intervals $I_j \subseteq \R$, we say that $B = \prod_{j=1}^m I_j \subseteq \R^m$ is a \emph{rectangle}. The intervals $I_j$ with endpoints $u_j \leq v_j$ may be open, closed, or half-open, and may be empty or unbounded. For any $\delta > 0$, denote by $B^{-\delta} := \prod_{j=1}^m [u_j + \delta, v_j - \delta]$ the rectangle obtained from $B$ by moving each finite endpoint ``inwards'' by $\delta$. If $a > b$ we let $[a,b] = \emptyset$, and when $B = \emptyset$ we let $B^{-\delta} := \emptyset$. 
\end{definition}

We now give the corresponding uniqueness event.

\begin{definition}[Uniqueness event for rectangles]\label{def:rectangle-unique-event}
    For any $m, r \in \N$, let $\parallelfunc: \R^m \to \signs$ be any measurable function and sample $\Gsuff_1, \dots, \Gsuff_r \sim \stdnormal$ independently. For any rectangle $B = \prod_{j=1}^m I_j$, define $\mathcal{U}_{\parallelfunc, B}$ as the event that the vector $\solsuff \in \signs^r$ given by $\solsuff_i = \parallelfunc(\Gsuff_i)$ is the unique element of $\signs^r$ such that
    \begin{align}
        \frac{1}{\sqrt{r}}\sum_{i=1}^r \solsuff_i \Gsuff_i \in B.
    \end{align}
\end{definition}

\Cref{lem:rectangle-anti-uniqueness} gives an upper bound on the probability of $\mathcal{U}_{\parallelfunc, B}$.

\begin{lemma}[Upper bound on uniqueness for rectangles]\label{lem:rectangle-anti-uniqueness}
    Fix any integers $m, r \geq 3$, any measurable function $\parallelfunc: \R^m \to \signs$, and any rectangle $B \subseteq \R^m$. Sample $\Gsuff_1, \dots, \Gsuff_r \sim \stdnormal$ independently and define $\mathcal{U}_{\parallelfunc, B}$ as in Definition~\ref{def:rectangle-unique-event}. Then
    \begin{align}
        \Prob{\mathcal{U}_{\parallelfunc, B}} \leq O\left(\frac1m + \frac{(\log m)^{3/2} \log r + (\log m)^2}{\sqrt{r}} + \frac{\log m}{\sqrt{r}}\right), \label{line:rectangle-uniqueness-bound}
    \end{align}
    with the probability taken over $\Gsuff_1, \dots, \Gsuff_r$. The implicit constant is absolute, so the bound is uniform over $\parallelfunc$ and $B$.
\end{lemma}

We prove \Cref{lem:rectangle-anti-uniqueness} in Appendix~\ref{ap:lemmas-for-unique}. The proof follows that of \Cref{thm:anti-uniqueness}.
We prove in \Cref{lem:unique-to-rectangle-shell} that the score $\score$ of the baseline will lie in a thin ``shell'' near the boundary of $B$ with high probability. We then appeal to the Berry-Esseen theorem and Nazarov's inequality as before, with the main change being that we treat each finite endpoint of $B$ as a separate coordinate when applying Nazarov's inequality. Assuming \Cref{lem:rectangle-anti-uniqueness}, we can now prove our main result for general perceptrons.

\begin{proof}[Proof of \Cref{thm:general-intro}]
    We follow the proof of \Cref{thm:main-deterministic}. By \Cref{lem:randomized}, it suffices to prove our bound for every deterministic $\A \in \online_K$, as long as the bound does not depend on $\A$. Moreover, since $\mathcal{S}^h(G, K) \subseteq \mathcal{S}^{h'}(G, K)$ whenever $h' \leq h$, it suffices to prove the result with $h$ replaced by $\min\{h, \lfloor n^{1/4} \rfloor\}$. Therefore we assume that $\omega(\log^4 n) < h(n) \leq n^{1/4}$.
    
    Fix an integer $r$ to be specified later, and define the algorithm $\A^{(r)}$. Sample $G \sim \G^{m,n}$, let $t(n) = h(n) / log^4 n$, and define $a(n) := \lceil \sqrt{t(n)} \log^4 n\rceil$. Define the events
    \begin{align}
        \success &:= \{\A(G, K) \in \mathcal{S}^h(G, K)\} \\
        \match_r &:= \{\A(G, K) = \A^{(r)}(G, K)\}.
    \end{align}
    \Cref{thm:successful-A-matches-baseline}, applied with $\mathcal{S}^h(G, K)$ in place of $\isosolset$, states that there exists some $r$ satisfying $a \leq r < h$ such that
    \begin{align}
        \Prob{\success} &= \Prob{\success \cap \match_r} + \Prob{\success \cap \match_r^c} \\
        &< \Prob{\success \cap \match_r} + o(1).
    \end{align}
    We now analyze the event $\success \cap \match_r$. Denote by $\Gpref$ the first $n - r$ columns of $G$ and denote the last $r$ columns by $\Gsuff_1, \dots, \Gsuff_r$. Write $\A(G, K) = \sigma' \| \solsuff$ for some $\solsuff \in \signs^r$. In the binary perceptron, the feasible suffix scores form an upper orthant. Here, each row score may lie in any of the intervals forming $K$, so the feasible suffix scores need not form a rectangle. However, once only $r \leq n^{1/4}$ columns remain, the row scores can only move by a small amount. Define the event
    \begin{align}
        \mathcal{L}_n := \left\{\max_{i \in [r],\, j \in [m]} |\Gsuff_{ij}| < 2\sqrt{\log n}\right\}.
    \end{align}
    By a union bound and a tail bound on standard Gaussian random variables, we have
    \begin{align}
        \Prob{\mathcal{L}_n^c} \leq 2mr\Exp{-2\log n} = \frac{2mr}{n^2} < o(1),
    \end{align}
    since $r < h \leq n^{1/4}$. Define $p := \Gpref\sigma'/\sqrt{n} \in \R^m$. On the event $\mathcal{L}_n$, for every $\solsuff' \in \signs^r$ and every $j \in [m]$ we have
    \begin{align}
        \left|\frac{\langle G_j, \sigma' \| \solsuff' \rangle}{\sqrt{n}} - p_j\right| = \frac{1}{\sqrt{n}}\left|\sum_{i=1}^r \solsuff'_i \Gsuff_{ij}\right| < \eps_n := \frac{2r\sqrt{\log n}}{\sqrt{n}}.
    \end{align}
    Since $r \leq n^{1/4}$, we have $\eps_n \to 0$, so for large enough $n$ we have $2\eps_n < \Delta$. Since $K$ is well-separated, the interval $(p_j - \eps_n, p_j + \eps_n)$ meets at most one of the intervals forming $K$. Let $I_j$ be that interval, or $I_j := \varnothing$ if there is none, and define the rectangle
    \begin{align}
        B := \prod_{j=1}^m \sqrt{\frac{n}{r}}\left(I_j - p_j\right),
    \end{align}
    which does not depend on $\Gsuff_1, \dots, \Gsuff_r$. Then on the event $\mathcal{L}_n$, for every $\solsuff' \in \signs^r$ we have
    \begin{align}
        \sigma' \| \solsuff' \in \mathcal{S}(G, K) \iff \frac{1}{\sqrt{r}}\sum_{i=1}^r \solsuff'_i \Gsuff_i \in B. \label{line:localized-rectangle}
    \end{align}
    We claim that $\success \cap \match_r \cap \mathcal{L}_n \subseteq \mathcal{U}_{\nextcol, B}$. To see why, note that $\success$ and the fact that $r < h$ together imply that $\solsuff$ is the unique string in $\signs^r$ such that $\sigma' \| \solsuff \in \mathcal{S}(G, K)$. By \Cref{line:localized-rectangle}, on the event $\mathcal{L}_n$ this means that $\solsuff$ is the unique string in $\signs^r$ that satisfies $r^{-1/2}\sum_{i=1}^r \solsuff_i \Gsuff_i \in B$. Let $\nextcol := \subA_{n-r+1}$, so $\match_r$ implies that $\nextcol(\Gsuff_i) = z_i$ for each $i \in [r]$. Conditional on $\Gpref$, the columns $\Gsuff_1, \dots, \Gsuff_r$ are independent with distribution $\stdnormal$, while $\nextcol$ and $B$ are fixed. Note that we use $\mathcal{L}_n$ only to identify the feasible suffixes with the rectangle $B$, and we do not condition on it, since that would change the distribution of the suffix columns. Therefore we have
    \begin{align}
        \Prob{\success \cap \match_r} &\leq \Prob{\mathcal{U}_{\nextcol, B}} + \Prob{\mathcal{L}_n^c} \\
        &\leq O\left(\frac1m + \frac{(\log m)^{3/2} \log r + (\log m)^2}{\sqrt{r}} + \frac{\log m}{\sqrt{r}}\right) + o(1) \tag{\Cref{lem:rectangle-anti-uniqueness}} \\
        &< o(1),
    \end{align}
    with the last inequality due to $m > \omega(1)$ and $r \geq a(n) > \omega(\log^4 m)$. As at the end of the proof of \Cref{thm:main-deterministic}, these bounds hold for every $r \geq a(n)$ and otherwise do not depend on $\A$, so the result follows.
\end{proof}

\section*{AI methodology}
The authors used AI models extensively throughout this project, starting with GPT-5.4 Pro and continuing through later versions up to GPT-6 Astra. The idea of the suffix-parallel baseline (\Cref{def:next-col-rule}) grew out of these interactions. An AI model then suggested a proof technique similar to the one presented in \Cref{sec:technical-overview,sec:proofs}, except that it only concerned linear isolation, and applied only to the binary perceptron. The authors then digested that technique, simplified it, and worked with an AI model to produce the proofs of \Cref{thm:main,thm:general-intro}, to extend the results to polylogarithmic isolation and general perceptrons. The authors also made limited use of large language models, including ChatGPT and Claude, to copyedit the writing, prepare figures, and check references. The authors take full responsibility for the mathematical content and the correctness of this paper.

\section*{Acknowledgments}
G.S.\ thanks Aloni Cohen for insightful discussions and for proofreading an earlier draft of this paper.

\appendix

\section{Additional proofs}\label{sec:additional-proofs}

\subsection{Auxiliary probability facts}

\newcommand{\eigenmin}{\lambda_{\text{min}}}

\begin{theorem}[Berry-Esseen \cite{chernozhukov2023nearly}]\label{thm:berry-esseen}
    For any integers $r, m \geq 3$, let $W_1, \dots, W_r \in \R^m$ be independent, identically distributed centered random vectors. Define
    \begin{align}
        W := \frac{1}{\sqrt{r}}\sum_{i=1}^r W_i
    \end{align}
    and let $\Sigma \in \R^{m\times m}$ be the covariance matrix of $W$. For each $j \in [m]$, write $\sigma_j^2 := \Sigma_{jj}$ and suppose that $0 < \sigma_j^2 < \infty$. Let $D \in \R^{m\times m}$ be the invertible, diagonal matrix with entries $(\sigma_1, \dots, \sigma_m)$ on the diagonal, and let $R := D^{-1}\Sigma D^{-1}$ be the correlation matrix of $W$, with smallest eigenvalue denoted by $\eigenmin(R)$. Suppose that there exist universal constants $K \geq 1$ and $b > 0$ such that the following statements hold:
    \begin{enumerate}
        \item $\subgaussian{W_{1j}/\sigma_j} \leq K$ for all $j \in [m]$.
        \item $\eigenmin(R) \geq b$.
    \end{enumerate}
    Sample $Z \sim \mathcal{N}(0, \Sigma)$. Then we have 
    \begin{align}
        \sup_{A \in \rectangleset} \left|\Prob{W \in A} - \Prob{Z \in A}\right| \leq O\left(\frac{(\log m)^{3/2}\log r + (\log m)^2}{\sqrt{r}}\right),
    \end{align}
    where
    \begin{align}
        \rectangleset = \{\Pi_{j=1}^m [a_j, b_j): -\infty \leq a_j \leq b_j \leq \infty \text{ for every } j \in [m]\}
    \end{align}
    is the set of rectangles in $\R^m$.
\end{theorem}

\begin{proof}
    The proof is a direct application of \cite[Corollary 2.1]{chernozhukov2023nearly} under conditions (E.2) and (M). Observe that $\Sigma = \Cov(W_1)$ because the $W_i$ are independent and identically distributed. For all $i \in [r]$, $j \in [m]$, we have $\subgaussian{W_{ij}/\sigma_j} \leq K$. By the definition of the subgaussian norm and the inequality $e^{x^2} \geq x^4 / 2$, we have
    \begin{align}
        \Ex{|W_{ij}/\sigma_j|^4} \leq 2K^4\Ex{\exp{\left(\frac{(W_{ij}/\sigma_j)^2}{K^2}\right)}} \leq 4K^4.
    \end{align}
    Therefore conditions (E.2) and (M) of \cite{chernozhukov2023nearly} hold with $n = r$, $d = m$, $X_i = W_i$, and $B_n = 2K^2 \geq K$. Moreover, the parameter $\sigma_{\star, W}$ of \cite{chernozhukov2023nearly} is the square root of the smallest eigenvalue of the correlation matrix of $W$, so $\sigma_{\star, W} = \sqrt{\eigenmin(R)} \geq \sqrt{b}$. The result follows from \cite[Corollary 2.1]{chernozhukov2023nearly}. That result is stated for one type of rectangle, and it extends to all of $\rectangleset$ by dominated convergence, since the indicator of any rectangle is a pointwise limit of indicators of rectangles of that type.
\end{proof}

\begin{theorem}[Nazarov's inequality \cite{nazarov2004maximal,chernozhukov2017detailedproofnazarovsinequality}]\label{thm:nazarov}
    Let $Z \in \R^m$ be an $m$-dimensional Gaussian random variable with mean $\mu \in \R^m$ and suppose there exists a constant $\varW > 0$ such that $\Var(Z_j) \geq b_0$ for all $j \in [m]$. Then for any $u \in \R^m$ and $\delta > 0$, we have
    \begin{align}
       \Prob{Z \in \orthant{u} \setminus \orthant{u + \delta \mathbf{1}_m}} \leq \frac{\delta}{\sqrt{\varW}}\left(\sqrt{2\log m} + 2\right)
    \end{align}
\end{theorem}
\begin{proof}
    The proof is a direct application of Nazarov's inequality (stated in the form of \cite[Theorem 1]{chernozhukov2017detailedproofnazarovsinequality}), which concerns the probability of centered Gaussian vectors lying in the difference of lower orthants of an $m$-dimensional space. Since our event of interest concerns upper orthants, we have to apply Nazarov's inequality to the centered, flipped version of $Z$. We define
    \begin{align}
        Y := -(Z - \mu)
    \end{align}
    and observe that for all $j \in [m]$ we have
    \begin{align}
        \Ex{Y_j} &= -(\Ex{Z_j} - \Ex{\mu_j}) = 0 \\
        \Ex{Y_j^2} &= \Ex{Y_j^2} - \Ex{Y_j}^2 = \Var(Y_j) = \Var(Z_j) \geq \varW,
    \end{align}
    so the hypotheses of \cite[Theorem 1]{chernozhukov2017detailedproofnazarovsinequality} hold. Fix any $u \in \R^m$ and $\delta > 0$, and choose $y := \Ex{Z} - u - \delta \mathbf{1}_m \in \R^m$. Then we have
    \begin{align}
        \Prob{Z \in \orthant{u} \setminus \orthant{u + \delta \mathbf{1}_m}} &= \Prob{Z \geq u} - \Prob{Z \geq u + \delta \mathbf{1}_m} \\
        &= \Prob{Y \leq y + \delta \mathbf{1}_m} - \Prob{Y \leq y} \\
        &\leq \frac{\delta}{\sqrt{\varW}}\left(\sqrt{2\log m} + 2\right) \tag*{\cite[Theorem 1]{chernozhukov2017detailedproofnazarovsinequality}}.
    \end{align}
\end{proof}

\subsection{Lemmas for upper bounds on uniqueness}\label{ap:lemmas-for-unique}

\begin{lemma}[Uniqueness implies thin shell]\label{lem:unique-to-orthant-switch}
    Fix any integers $m, r \geq 2$. Define $\Gsuff_1, \dots, \Gsuff_r$, $\parallelfunc$, $\residual$, and $\unique$ as in \Cref{def:unique-event}. Let
    \begin{align}
        S := \frac{1}{\sqrt{r}}\sum_{i=1}^r \parallelfunc(\Gsuff_i) \Gsuff_i.
    \end{align}
    Then there exists some $\delta = \Theta(\sqrt{(\log m)/r})$ such that
    \begin{align}
        \Prob{\unique} \leq \Prob{S \in \orthant{\residual} \setminus \orthant{\residual + \delta \mathbf{1}_m}} + O\left(\frac{1}{m}\right).
    \end{align}
\end{lemma}
\begin{proof}
    The event $\unique$ implies that for all $j \in [m]$ we have
    \begin{align}
        S_j \geq \residual_j. \label{line:S-feasible}
    \end{align}
    Moreover, the event also implies that swapping the sign for any term in the sum of $S$, say $i = 1$, ensures that there exists some $j \in [m]$ such that
    \begin{align}
        S_j - \frac{2\parallelfunc(\Gsuff_1)}{\sqrt{r}}\Gsuff_{1j} < \residual_j. \label{line:swapping-decreases}
    \end{align}
    Let $L_m := C \sqrt{\log m}$ for some $C > 2$, and condition on the event $\mathcal{E}_m := \{\max_{j \in [m]} |\Gsuff_{1j}| \leq L_m\}$. Combining \Cref{line:S-feasible,line:swapping-decreases} gives us
    \begin{align}
        0 &\leq S_j - \residual_j \\
        &< \frac{2\parallelfunc(\Gsuff_1)}{\sqrt{r}}\Gsuff_{1j} \\
        &\leq \frac{2|\Gsuff_{1j}|}{\sqrt{r}} \\
        &\leq \frac{2 L_m}{\sqrt{r}},
    \end{align}
    where the last inequality holds on the event $\mathcal{E}_m$. Therefore setting
    \begin{align}
        \delta := \frac{2L_m}{\sqrt{r}} = \Theta\left(\sqrt{\frac{\log m}{r}}\right)
    \end{align}
    we have
    \begin{align}
        \unique \cap \mathcal{E}_m \subseteq \{S \in \orthant{\residual} \setminus \orthant{\residual + \delta\mathbf{1}_m}\}.
    \end{align}
    By the law of total probability, we have
    \begin{align}
        \Prob{\unique} &= \Prob{\unique \cap \mathcal{E}_m} + \Prob{\unique  \cap \mathcal{E}_m^c} \\
        &\leq \Prob{S \in \orthant{\residual} \setminus \orthant{\residual + \delta\mathbf{1}_m}} + \Prob{\mathcal{E}_m^c}.
    \end{align}
    By a union bound and a tail bound on standard Gaussian random variables, we have
    \begin{align}
        \Prob{\mathcal{E}_m^c} &\leq 2m\Exp{-\frac{L_m^2}{2}} \\
        &\leq 2m^{1 - C^2/2} \\
        &< 2m^{-1},
    \end{align}
    and the result follows.
\end{proof}

\begin{lemma}[Uniqueness implies thin shell for rectangles]\label{lem:unique-to-rectangle-shell}
    Fix any integers $m, r \geq 2$. Define $\Gsuff_1, \dots, \Gsuff_r$, $\parallelfunc$, $B$, and $\mathcal{U}_{\parallelfunc, B}$ as in \Cref{def:rectangle-unique-event}. Let
    \begin{align}
        S := \frac{1}{\sqrt{r}}\sum_{i=1}^r \parallelfunc(\Gsuff_i) \Gsuff_i.
    \end{align}
    Then there exists some $\delta = \Theta(\sqrt{(\log m)/r})$ such that
    \begin{align}
        \Prob{\mathcal{U}_{\parallelfunc, B}} \leq \Prob{S \in B \setminus B^{-\delta}} + O\left(\frac{1}{m}\right).
    \end{align}
\end{lemma}
\begin{proof}
    The event $\mathcal{U}_{\parallelfunc, B}$ implies that $S \in B$. Moreover, the event also implies that swapping the sign of any term in the sum of $S$, say $i = 1$, moves $S$ outside of $B$, so there exists some $j \in [m]$ such that
    \begin{align}
        S_j - \frac{2\parallelfunc(\Gsuff_1)}{\sqrt{r}}\Gsuff_{1j} \notin I_j. \label{line:rectangle-swapping}
    \end{align}
    Let $L_m := C\sqrt{\log m}$ for some $C > 2$, and consider the event $\mathcal{E}_m := \{\max_{j \in [m]} |\Gsuff_{1j}| < L_m\}$. Setting
    \begin{align}
        \delta := \frac{2L_m}{\sqrt{r}} = \Theta\left(\sqrt{\frac{\log m}{r}}\right),
    \end{align}
    we have $|2\parallelfunc(\Gsuff_1)\Gsuff_{1j}/\sqrt{r}| < \delta$ for every $j \in [m]$ on the event $\mathcal{E}_m$. So if $S \in B^{-\delta}$, then for every $j \in [m]$ we would have
    \begin{align}
        S_j - \frac{2\parallelfunc(\Gsuff_1)}{\sqrt{r}}\Gsuff_{1j} \in (u_j, v_j) \subseteq I_j,
    \end{align}
    which contradicts \Cref{line:rectangle-swapping}. Therefore
    \begin{align}
        \mathcal{U}_{\parallelfunc, B} \cap \mathcal{E}_m \subseteq \{S \in B \setminus B^{-\delta}\}.
    \end{align}
    The result follows from the law of total probability and the bound $\Prob{\mathcal{E}_m^c} < 2m^{-1}$, exactly as in the proof of \Cref{lem:unique-to-orthant-switch}.
\end{proof}

\begin{lemma}[Hypotheses for Berry-Esseen hold]\label{lem:hypotheses-for-berry-esseen}
    Fix any integers $m, r \geq 3$. Let $\Gsuff_1, \dots, \Gsuff_r \sim \stdnormal$ independently and let $\parallelfunc: \R^m \to \signs$ be measurable. For each $i \in [r]$, define $S_i := \parallelfunc(\Gsuff_i) \Gsuff_i$ and $W_i := S_i - \Ex{S_i}$. Define
    \begin{align}
        W := \frac{1}{\sqrt{r}}\sum_{i=1}^r W_i
    \end{align}
    and let $\Sigma \in \R^{m \times m}$ be the covariance matrix of $W$. For each $j \in [m]$, write $\sigma_j^2 := \Sigma_{jj}$. The following statements hold.
    \begin{enumerate}
        \item The random variables $W_i$ are independent and identically distributed. \label{line:iid}
        \item $0 < \sigma_j^2 < \infty$ for all $j \in [m]$. \label{line:constant-variance}
    \end{enumerate}
    Let $D \in \R^{m \times m}$ be the invertible, diagonal matrix with entries $(\sigma_1, \dots, \sigma_m)$ on the diagonal, and let $R = D^{-1}\Sigma D^{-1}$ be the correlation matrix of $W$, with smallest eigenvalue denoted by $\eigenmin(R)$. Then there exist universal constants $K \geq 1$ and $b > 0$ such that the following statements hold:
    \begin{enumerate}[start=3]
        \item $\subgaussian{W_{1j}/\sigma_j} \leq K$ for all $j \in [m]$. \label{line:bounded-subgaussian-norm}
        \item $\eigenmin(R) \geq b$. \label{line:bounded-eigenvalue}
    \end{enumerate}
\end{lemma}
\begin{proof}
    \Cref{line:iid} follows because $\parallelfunc$ is fixed and the random variables $\Gsuff_i$ are independent and identically distributed. Let $\mu = \Ex{S_i}$, which is identical for all $i \in [r]$. We claim that
    \begin{align}
        \norm{\mu}_2 \leq \sqrt{\frac{2}{\pi}}.\label{line:mu-bound}
    \end{align}
    The claim is trivial when $\norm{\mu}_2 = 0$. Otherwise, consider the unit vector $u = \mu / \norm{\mu}_2$ and see that
    \begin{align}
        \norm{\mu}_2 &= |u^T \mu| \\
        &= \left|\Ex{\parallelfunc(\Gsuff_1)u^T \Gsuff_1}\right| \\
        &\leq \Ex{\left|u^T\Gsuff_1\right|} \\
        &= \sqrt{\frac{2}{\pi}}
    \end{align}
    with the last equality using the fact that $u^T\Gsuff_1 \sim \mathcal{N}(0,1)$. Using the fact that the $W_i$ are independent and identically distributed, we can write
    \begin{align}
        \Sigma &= \Cov(S_1) \\
        &= \Ex{S_1 S_1^T} - \mu \mu^T \\
        &= \Ex{\phi(\Gsuff_1)^2 \Gsuff_1 \Gsuff_1^T} - \mu \mu^T \\
        &= I_m - \mu \mu^T \label{line:cov-W}
    \end{align}
    Set $\varW := 1 - \frac{2}{\pi} > 0$. Recalling that $\sigma_j^2 := \Sigma_{jj}$, \Cref{line:mu-bound,line:cov-W} imply that
    \begin{align}
        \sqrt{\varW} \leq \sigma_j \leq 1, \label{line:bounds-on-sigma}
    \end{align}
    which proves \Cref{line:constant-variance}. Moreover, note that $|S_{ij}| = |Y_{ij}|$. Therefore the triangle inequality for the subgaussian norm, along with \Cref{line:mu-bound,line:bounds-on-sigma}, imply that there exists some absolute constant $K \geq 1$ such that
    \begin{align}
        \subgaussian{W_{1j}/\sigma_j} &= \frac{1}{\sigma_j}\subgaussian{S_{1j} - \mu_j} \\
        &\leq \frac{1}{\sigma_j} \left(\subgaussian{S_{1j}} + \subgaussian{\mu_j}\right) \\
        &\leq K,
    \end{align}
    which proves \Cref{line:bounded-subgaussian-norm}. Finally, recall that $R = D^{-1}\Sigma D^{-1}$, where $D$ is the diagonal matrix with $\sigma_1, \dots, \sigma_m$ on the diagonal. Every eigenvector $x \in \R^m$ of $R$ with eigenvalue $\lambda$ satisfies the equality $x^TRx = \lambda x^T x = \lambda \norm{x}_2^2$. Furthermore, Cauchy-Schwarz gives that for any $x \in \R^m$, we have
    \begin{align}
        x^T \Sigma x &= x^T(I_m - \mu \mu^T)x \\
        &= \norm{x}_2^2 - (x^T \mu)^2 \\
        &\geq \norm{x}_2^2 - \norm{x}_2^2 \norm{\mu}_2^2 \\
        &\geq \varW \norm{x}_2^2 \label{line:loewner-order}
    \end{align}
    Therefore we have
    \begin{align}
        x^T R x &= x^T D^{-1}\Sigma D^{-1} x \\
        &= (D^{-1}x)^T \Sigma (D^{-1}x) \\
        &\geq \varW \sum_{j=1}^m \left(\frac{x_j}{\sigma_j}\right)^2 \tag{\Cref{line:loewner-order}} \\
        &\geq \varW \norm{x}_2^2 \tag{\Cref{line:bounds-on-sigma}}
    \end{align}
    and so the smallest eigenvalue of $R$ is $\eigenmin(R) \geq \varW > 0$, proving \Cref{line:bounded-eigenvalue}.
\end{proof}

\begin{proof}[Proof of \Cref{lem:rectangle-anti-uniqueness}]
    We follow the proof of \Cref{thm:anti-uniqueness}. If $B = \varnothing$, then $\mathcal{U}_{\parallelfunc, B} = \varnothing$ and there is nothing to prove, so we assume that $B \neq \varnothing$. For clarity, define $S_i := \parallelfunc(\Gsuff_i) \Gsuff_i$ for each $i \in [r]$ and let
    \begin{align}
        S := \frac{1}{\sqrt{r}} \sum_{i = 1}^r S_i.
    \end{align}
    By \Cref{lem:unique-to-rectangle-shell}, there exists a $\delta = \Theta(\sqrt{(\log m)/r})$ such that
    \begin{align}
        \Prob{\mathcal{U}_{\parallelfunc, B}} \leq \Prob{S \in B \setminus B^{-\delta}} + O\left(\frac{1}{m}\right). \label{line:rectangle-thin-shell}
    \end{align}
    We now proceed to bound $\Prob{S \in B \setminus B^{-\delta}}$. Define $\mu$, $W$, $\Sigma$, $Z \sim \mathcal{N}(0, \Sigma)$, and $\rho_{m,r}$ as in the proof of \Cref{thm:anti-uniqueness}, so that $\rho_{m,r}$ satisfies the bound in \Cref{line:berry-esseen}. The same bound holds with the supremum taken over all rectangles, since the indicator of every rectangle is a pointwise limit of indicators of rectangles in $\rectangleset$, and both probabilities converge by dominated convergence. Since $W = S - \sqrt{r}\mu$ and $B^{-\delta} \subseteq B$, the same argument as in the proof of \Cref{thm:anti-uniqueness} gives
    \begin{align}
        \Prob{S \in B \setminus B^{-\delta}} &= \Prob{W \in (B - \sqrt{r}\mu) \setminus (B^{-\delta} - \sqrt{r}\mu)} \\
        &\leq \Prob{Z + \sqrt{r}\mu \in B \setminus B^{-\delta}} + 2\rho_{m,r}. \label{line:rectangle-gaussian-comparison}
    \end{align}
    Finally, we bound the Gaussian probability in \Cref{line:rectangle-gaussian-comparison} using Nazarov's inequality (\Cref{thm:nazarov}). \Cref{thm:nazarov} concerns upper orthants, while $B$ may have two finite endpoints in each coordinate, so we treat each finite endpoint as a separate coordinate. Let $J_- := \{j \in [m]: u_j > -\infty\}$ and $J_+ := \{j \in [m]: v_j < \infty\}$, and let $m' := |J_-| + |J_+| \leq 2m$. If $m' = 0$, then $B = B^{-\delta} = \R^m$ and there is nothing to prove. Otherwise, define the Gaussian random variable $V \in \R^{m'}$ and the vector $q \in \R^{m'}$ by
    \begin{align}
        V := \left((Z_j + \sqrt{r}\mu_j)_{j \in J_-},\, (-Z_j - \sqrt{r}\mu_j)_{j \in J_+}\right), \qquad q := \left((u_j)_{j \in J_-},\, (-v_j)_{j \in J_+}\right).
    \end{align}
    Then $\{Z + \sqrt{r}\mu \in B^{-\delta}\} = \{V \in \orthant{q + \delta\mathbf{1}_{m'}}\}$, while the events $\{Z + \sqrt{r}\mu \in B\}$ and $\{V \in \orthant{q}\}$ differ only by an event of probability zero, since each $Z_j$ has a continuous distribution. By \Cref{lem:hypotheses-for-berry-esseen}, we have $\Sigma_{jj} \geq \varW > 0$ for each $j \in [m]$, so $\Var(V_k) \geq \varW$ for each $k \in [m']$. Therefore \Cref{thm:nazarov} gives
    \begin{align}
        \Prob{Z + \sqrt{r}\mu \in B \setminus B^{-\delta}} &= \Prob{V \in \orthant{q} \setminus \orthant{q + \delta\mathbf{1}_{m'}}} \\
        &\leq \frac{\delta}{\sqrt{\varW}}\left(\sqrt{2\log (2m)} + 2\right). \label{line:rectangle-nazarov}
    \end{align}
    Note that \Cref{thm:nazarov} only requires a lower bound on the variance of each coordinate, so it applies even though $V$ may have a singular covariance matrix. Putting together \Cref{line:rectangle-thin-shell,line:rectangle-gaussian-comparison,line:rectangle-nazarov}, we have
    \begin{align}
        \Prob{\mathcal{U}_{\parallelfunc, B}} \leq O\left(\frac{1}{m}\right) + O\left(\frac{(\log m)^{3/2}\log r + (\log m)^2}{\sqrt{r}}\right) + \frac{\delta}{\sqrt{\varW}}\left(\sqrt{2\log (2m)} + 2\right).
    \end{align}
    Recalling that $\delta = \Theta(\sqrt{(\log m)/r})$ and $\varW$ is a positive constant completes the proof.
\end{proof}

\subsection{Reduction to deterministic algorithms}\label{pf:sec:setup}

\begin{lemma}\label{lem:randomized}
    Let $\online(n)$ be the set of algorithms in $\online$ that accept input matrices with $n$ columns. Let $b_n$ be an upper bound on the success probability of all deterministic algorithms $\A^* \in \online(n)$. Then $b_n$ is an upper bound on the success probability of any algorithm $\A \in \online(n)$.
\end{lemma}
\begin{proof}
    Suppose for contradiction that there exists a randomized algorithm $\A \in \online(n)$ that succeeds with probability strictly greater than $b_n$. By the law of total expectation, we can write
    \begin{align}
        \mathbb{E}_\xi\left[\mathbb{E}_G\left[\mathbf{1}_{\{\A(G, \margin; \xi) \in \isosolset\}} \mid \xi \right] \right] > b_n,
    \end{align}
    where $\xi$ denotes the internal randomness of $\A$. By an averaging argument, there must exist a random seed $\xi^*$ such that
    \begin{align}
        \mathbb{E}_G \left[\mathbf{1}_{\{\A(G, \margin; \xi^*) \in \isosolset\}}\right] > b_n.
    \end{align}
    This implies that the deterministic algorithm $\A^*(G, \margin) := \A(G, \margin; \xi^*)$ succeeds with probability strictly greater than $b_n$, which is a contradiction.
\end{proof}

\bibliographystyle{alpha}
\bibliography{ref}

@inproceedings{nazarov2004maximal,
  title={On the maximal perimeter of a convex set in with respect to a Gaussian measure},
  author={Nazarov, Fedor},
  booktitle={Geometric Aspects of Functional Analysis: Israel Seminar 2001-2002},
  pages={169--187},
  year={2004},
  organization={Springer}
}

@article{chernozhukov2023nearly,
  author  = {Chernozhukov, Victor and Chetverikov, Denis and Koike, Yuta},
  title   = {Nearly optimal central limit theorem and bootstrap approximations in high dimensions},
  journal = {The Annals of Applied Probability},
  volume  = {33},
  number  = {3},
  pages   = {2374--2425},
  year    = {2023},
}

@misc{chernozhukov2017detailedproofnazarovsinequality,
  author       = {Chernozhukov, Victor and Chetverikov, Denis and Kato, Kengo},
  title        = {Detailed proof of {Nazarov}'s inequality},
  howpublished = {Preprint, \href{https://arxiv.org/abs/1711.10696}{arXiv:1711.10696}},
  year         = {2017},
}

@article{Cover1965,
  author  = {Cover, Thomas M.},
  title   = {Geometrical and Statistical Properties of Systems of Linear Inequalities with Applications in Pattern Recognition},
  journal = {IEEE Transactions on Electronic Computers},
  volume  = {EC-14},
  number  = {3},
  pages   = {326--334},
  year    = {1965},
}

@article{Gardner1988,
  author  = {Gardner, Elizabeth},
  title   = {The Space of Interactions in Neural Network Models},
  journal = {Journal of Physics A: Mathematical and General},
  volume  = {21},
  number  = {1},
  pages   = {257--270},
  year    = {1988},
}

@article{GardnerDerrida1988,
  author  = {Gardner, Elizabeth and Derrida, Bernard},
  title   = {Optimal Storage Properties of Neural Network Models},
  journal = {Journal of Physics A: Mathematical and General},
  volume  = {21},
  number  = {1},
  pages   = {271--284},
  year    = {1988},
}

@article{KrauthMezard1989,
  author  = {Krauth, Werner and M{\'e}zard, Marc},
  title   = {Storage Capacity of Memory Networks with Binary Couplings},
  journal = {Journal de Physique},
  volume  = {50},
  number  = {20},
  pages   = {3057--3066},
  year    = {1989},
}

@article{KimRoche1998,
  author  = {Kim, Jeong Han and Roche, James R.},
  title   = {Covering Cubes by Random Half Cubes, with Applications to Binary Neural Networks},
  journal = {Journal of Computer and System Sciences},
  volume  = {56},
  number  = {2},
  pages   = {223--252},
  year    = {1998},
}

@inproceedings{DingSun2019,
  author    = {Ding, Jian and Sun, Nike},
  title     = {Capacity Lower Bound for the {Ising} Perceptron},
  booktitle = {Proceedings of the 51st Annual ACM SIGACT Symposium on Theory of Computing (STOC)},
  pages     = {816--827},
  year      = {2019},
}

@article{Xu2021,
  author  = {Xu, Changji},
  title   = {Sharp Threshold for the {Ising} Perceptron Model},
  journal = {The Annals of Probability},
  volume  = {49},
  number  = {5},
  pages   = {2399--2415},
  year    = {2021},
}

@inproceedings{NakajimaSun2023,
  author    = {Nakajima, Shuta and Sun, Nike},
  title     = {Sharp Threshold Sequence and Universality for {Ising} Perceptron Models},
  booktitle = {Proceedings of the 2023 Annual ACM-SIAM Symposium on Discrete Algorithms (SODA)},
  pages     = {638--674},
  year      = {2023},
}

@inproceedings{LiSchrammZhou2025,
  author    = {Li, Shuangping and Schramm, Tselil and Zhou, Kangjie},
  title     = {Discrepancy Algorithms for the Binary Perceptron},
  booktitle = {Proceedings of the 57th Annual ACM Symposium on Theory of Computing (STOC)},
  pages     = {1668--1679},
  year      = {2025},
}

@article{BraunsteinZecchina2006,
  author  = {Braunstein, Alfredo and Zecchina, Riccardo},
  title   = {Learning by Message Passing in Networks of Discrete Synapses},
  journal = {Physical Review Letters},
  volume  = {96},
  number  = {3},
  pages   = {030201},
  year    = {2006},
}

@article{BaldassiEtAl2015,
  author  = {Baldassi, Carlo and Ingrosso, Alessandro and Lucibello, Carlo and Saglietti, Luca and Zecchina, Riccardo},
  title   = {Subdominant Dense Clusters Allow for Simple Learning and High Computational Performance in Neural Networks with Discrete Synapses},
  journal = {Physical Review Letters},
  volume  = {115},
  pages   = {128101},
  year    = {2015},
}

@article{BaldassiEtAl2020,
  author  = {Baldassi, Carlo and Della Vecchia, Riccardo and Lucibello, Carlo and Zecchina, Riccardo},
  title   = {Clustering of Solutions in the Symmetric Binary Perceptron},
  journal = {Journal of Statistical Mechanics: Theory and Experiment},
  volume  = {2020},
  number  = {7},
  pages   = {073303},
  year    = {2020},
}

@article{HuangKabashima2014,
  author  = {Huang, Haiping and Kabashima, Yoshiyuki},
  title   = {Origin of the Computational Hardness for Learning with Binary Synapses},
  journal = {Physical Review E},
  volume  = {90},
  pages   = {052813},
  year    = {2014},
}

@article{AubinPerkinsZdeborova2019,
  author  = {Aubin, Benjamin and Perkins, Will and Zdeborov{\'a}, Lenka},
  title   = {Storage Capacity in Symmetric Binary Perceptrons},
  journal = {Journal of Physics A: Mathematical and Theoretical},
  volume  = {52},
  number  = {29},
  pages   = {294003},
  year    = {2019},
}

@inproceedings{PerkinsXu2021,
  author    = {Perkins, Will and Xu, Changji},
  title     = {Frozen 1-{RSB} Structure of the Symmetric {Ising} Perceptron},
  booktitle = {Proceedings of the 53rd Annual ACM SIGACT Symposium on Theory of Computing (STOC)},
  pages     = {1579--1588},
  year      = {2021},
}

@inproceedings{AbbeLiSly2022Contiguity,
  author    = {Abbe, Emmanuel and Li, Shuangping and Sly, Allan},
  title     = {Proof of the Contiguity Conjecture and Lognormal Limit for the Symmetric Perceptron},
  booktitle = {2021 IEEE 62nd Annual Symposium on Foundations of Computer Science (FOCS)},
  pages     = {327--338},
  year      = {2021},
}

@inproceedings{AbbeLiSly2022RareCluster,
  author    = {Abbe, Emmanuel and Li, Shuangping and Sly, Allan},
  title     = {Binary Perceptron: Efficient Algorithms Can Find Solutions in a Rare Well-Connected Cluster},
  booktitle = {Proceedings of the 54th Annual ACM SIGACT Symposium on Theory of Computing (STOC)},
  pages     = {860--873},
  year      = {2022},
}

@inproceedings{GongHuangLiSellke2026,
  author    = {Gong, Shuyang and Huang, Brice and Li, Shuangping and Sellke, Mark},
  title     = {Stable Algorithms Cannot Reliably Find Isolated Perceptron Solutions},
  booktitle = {2026 IEEE 67th Annual Symposium on Foundations of Computer Science (FOCS)},
  year      = {2026},
}

@article{Pitt1982,
  author  = {Pitt, Loren D.},
  title   = {Positively Correlated Normal Variables Are Associated},
  journal = {The Annals of Probability},
  volume  = {10},
  number  = {2},
  pages   = {496--499},
  year    = {1982},
}

@article{Gamarnik2021,
  author  = {Gamarnik, David},
  title   = {The Overlap Gap Property: A Topological Barrier to Optimizing over Random Structures},
  journal = {Proceedings of the National Academy of Sciences},
  volume  = {118},
  number  = {41},
  pages   = {e2108492118},
  year    = {2021},
}

@inproceedings{GamarnikJagannathWein2020,
  author    = {Gamarnik, David and Jagannath, Aukosh and Wein, Alexander S.},
  title     = {Low-Degree Hardness of Random Optimization Problems},
  booktitle = {2020 IEEE 61st Annual Symposium on Foundations of Computer Science (FOCS)},
  pages     = {131--140},
  year      = {2020},
}

@inproceedings{GamarnikKizildagPerkinsXu2022,
  author    = {Gamarnik, David and K{\i}z{\i}lda{\u{g}}, Eren C. and Perkins, Will and Xu, Changji},
  title     = {Algorithms and Barriers in the Symmetric Binary Perceptron Model},
  booktitle = {2022 IEEE 63rd Annual Symposium on Foundations of Computer Science (FOCS)},
  pages     = {576--587},
  year      = {2022},
}

@inproceedings{GamarnikKizildagPerkinsXu2023,
  author    = {Gamarnik, David and K{\i}z{\i}lda{\u{g}}, Eren C. and Perkins, Will and Xu, Changji},
  title     = {Geometric Barriers for Stable and Online Algorithms for Discrepancy Minimization},
  booktitle = {Proceedings of the 36th Conference on Learning Theory (COLT)},
  pages     = {3231--3263},
  year      = {2023},
}

@article{BansalSpencer2020,
  author  = {Bansal, Nikhil and Spencer, Joel H.},
  title   = {On-Line Balancing of Random Inputs},
  journal = {Random Structures \& Algorithms},
  volume  = {57},
  number  = {4},
  pages   = {879--891},
  year    = {2020},
}

@article{BhamidiGamarnikGong2026,
  author  = {Bhamidi, Shankar and Gamarnik, David and Gong, Shuyang},
  title   = {Finding a Dense Submatrix of a Random Matrix. {S}harp Bounds for Online Algorithms},
  journal = {Electronic Communications in Probability},
  volume  = {31},
  pages   = {1--12},
  year    = {2026},
}

@misc{DhawanKizildagMaitra2025,
  author       = {Dhawan, Abhishek and K{\i}z{\i}lda{\u{g}}, Eren C. and Maitra, Neeladri},
  title        = {Sharp Online Hardness for Large Balanced Independent Sets},
  howpublished = {Preprint, \href{https://arxiv.org/abs/2508.20785}{arXiv:2508.20785}},
  year         = {2025},
}

@misc{GongLi2026,
  author       = {Gong, Shuyang and Li, Shuangping},
  title        = {Strong freezing in binary perceptron models},
  howpublished = {Preprint, \href{https://arxiv.org/abs/2610.06275}{arXiv:2610.06275}},
  year         = {2026},
}

@inproceedings{BJSS20,
  author    = {Bansal, Nikhil and Jiang, Haotian and Singla, Sahil and Sinha, Makrand},
  title     = {Online vector balancing and geometric discrepancy},
  booktitle = {Proceedings of the 52nd Annual ACM SIGACT Symposium on Theory of Computing (STOC)},
  pages     = {1139--1152},
  year      = {2020},
}

@inproceedings{KRR24,
  author    = {Kulkarni, Janardhan and Reis, Victor and Rothvoss, Thomas},
  title     = {Optimal online discrepancy minimization},
  booktitle = {Proceedings of the 56th Annual ACM Symposium on Theory of Computing (STOC)},
  pages     = {1832--1840},
  year      = {2024},
}

@article{Ros58,
  author  = {Rosenblatt, Frank},
  title   = {The perceptron: a probabilistic model for information storage and organization in the brain},
  journal = {Psychological Review},
  volume  = {65},
  number  = {6},
  pages   = {386--408},
  year    = {1958},
}

@article{HWK13,
  author  = {Huang, Haiping and Wong, K. Y. Michael and Kabashima, Yoshiyuki},
  title   = {Entropy landscape of solutions in the binary perceptron problem},
  journal = {Journal of Physics A: Mathematical and Theoretical},
  volume  = {46},
  number  = {37},
  pages   = {375002},
  year    = {2013},
}

@article{BBBZ07,
  author  = {Baldassi, Carlo and Braunstein, Alfredo and Brunel, Nicolas and Zecchina, Riccardo},
  title   = {Efficient supervised learning in networks with binary synapses},
  journal = {Proceedings of the National Academy of Sciences},
  volume  = {104},
  number  = {26},
  pages   = {11079--11084},
  year    = {2007},
}

@article{BBCILSZ16,
  author  = {Baldassi, Carlo and Borgs, Christian and Chayes, Jennifer T. and Ingrosso, Alessandro and Lucibello, Carlo and Saglietti, Luca and Zecchina, Riccardo},
  title   = {Unreasonable effectiveness of learning neural networks: from accessible states and robust ensembles to basic algorithmic schemes},
  journal = {Proceedings of the National Academy of Sciences},
  volume  = {113},
  number  = {48},
  pages   = {E7655--E7662},
  year    = {2016},
}

@article{Bal09,
  author  = {Baldassi, Carlo},
  title   = {Generalization learning in a perceptron with binary synapses},
  journal = {Journal of Statistical Physics},
  volume  = {136},
  number  = {5},
  pages   = {902--916},
  year    = {2009},
}

@article{BB15,
  author  = {Baldassi, Carlo and Braunstein, Alfredo},
  title   = {A max-sum algorithm for training discrete neural networks},
  journal = {Journal of Statistical Mechanics: Theory and Experiment},
  volume  = {2015},
  number  = {8},
  pages   = {P08008},
  year    = {2015},
}

@inproceedings{Hua24,
  author    = {Huang, Brice},
  title     = {Capacity threshold for the {Ising} perceptron},
  booktitle = {2024 IEEE 65th Annual Symposium on Foundations of Computer Science (FOCS)},
  pages     = {1126--1136},
  year      = {2024},
}

@misc{Shm26,
  author       = {Shmalo, Yitzchak},
  title        = {The storage capacity of the {Ising} perceptron: verification of the outstanding numerical conditions},
  howpublished = {Preprint, \href{https://arxiv.org/abs/2607.11010}{arXiv:2607.11010}},
  year         = {2026},
}

@article{Tal99,
  author  = {Talagrand, Michel},
  title   = {Intersecting random half cubes},
  journal = {Random Structures \& Algorithms},
  volume  = {15},
  number  = {3--4},
  pages   = {436--449},
  year    = {1999},
}

@article{Spe85,
  author  = {Spencer, Joel},
  title   = {Six standard deviations suffice},
  journal = {Transactions of the American Mathematical Society},
  volume  = {289},
  number  = {2},
  pages   = {679--706},
  year    = {1985},
}

@inproceedings{Ban10,
  author    = {Bansal, Nikhil},
  title     = {Constructive algorithms for discrepancy minimization},
  booktitle = {2010 IEEE 51st Annual Symposium on Foundations of Computer Science (FOCS)},
  pages     = {3--10},
  year      = {2010},
}

@inproceedings{LM15,
  author    = {Lovett, Shachar and Meka, Raghu},
  title     = {Constructive discrepancy minimization by walking on the edges},
  booktitle = {2012 IEEE 53rd Annual Symposium on Foundations of Computer Science (FOCS)},
  pages     = {61--67},
  year      = {2012},
}

@inproceedings{ALS21b,
  author    = {Alweiss, Ryan and Liu, Yang P. and Sawhney, Mehtaab},
  title     = {Discrepancy minimization via a self-balancing walk},
  booktitle = {Proceedings of the 53rd Annual ACM SIGACT Symposium on Theory of Computing (STOC)},
  pages     = {14--20},
  year      = {2021},
}

@inproceedings{VV25,
  author    = {Vafa, Neekon and Vaikuntanathan, Vinod},
  title     = {Symmetric perceptrons, number partitioning and lattices},
  booktitle = {Proceedings of the 57th Annual ACM Symposium on Theory of Computing (STOC)},
  pages     = {2191--2202},
  year      = {2025},
}

@article{AG25,
  author  = {{El Alaoui}, Ahmed and Gamarnik, David},
  title   = {Hardness of sampling solutions from the symmetric binary perceptron},
  journal = {Random Structures \& Algorithms},
  volume  = {66},
  number  = {4},
  pages   = {e70021},
  year    = {2025},
}

@misc{JL26,
  author       = {Jo, Sunghyeon and Lee, Taekyun},
  title        = {The exact online threshold for the asymmetric binary perceptron},
  howpublished = {Preprint, \href{https://arxiv.org/abs/2609.02124}{arXiv:2609.02124}},
  year         = {2026},
}

@inproceedings{AC08,
  author    = {Achlioptas, Dimitris and Coja-Oghlan, Amin},
  title     = {Algorithmic barriers from phase transitions},
  booktitle = {2008 49th Annual IEEE Symposium on Foundations of Computer Science (FOCS)},
  pages     = {793--802},
  year      = {2008},
}

@article{ZK07,
  author  = {Zdeborov{\'a}, Lenka and Krzakala, Florent},
  title   = {Phase transitions in the coloring of random graphs},
  journal = {Physical Review E},
  volume  = {76},
  number  = {3},
  pages   = {031131},
  year    = {2007},
}

@inproceedings{Mol18,
  author    = {Molloy, Michael},
  title     = {The freezing threshold for {$k$}-colourings of a random graph},
  booktitle = {Proceedings of the 44th Annual ACM Symposium on Theory of Computing (STOC)},
  pages     = {921--930},
  year      = {2012},
}

@article{BAKZ24,
  author  = {Barbier, Damien and {El Alaoui}, Ahmed and Krzakala, Florent and Zdeborov{\'a}, Lenka},
  title   = {On the atypical solutions of the symmetric binary perceptron},
  journal = {Journal of Physics A: Mathematical and Theoretical},
  volume  = {57},
  number  = {19},
  pages   = {195202},
  year    = {2024},
}

@misc{HSS26,
  author       = {Huang, Brice and Sellke, Mark and Sun, Nike},
  title        = {Algorithmic threshold for high-dimensional projection pursuit {I}: general theory},
  howpublished = {Preprint, \href{https://arxiv.org/abs/2608.29416}{arXiv:2608.29416}},
  year         = {2026},
}

@misc{FJLN26,
  author       = {Fiedler, Christian and Jackson, Joe and Lacker, Daniel and Niles-Weed, Jonathan},
  title        = {The mean-field limit of online stochastic vector balancing},
  howpublished = {Preprint, \href{https://arxiv.org/abs/2605.14149}{arXiv:2605.14149}},
  year         = {2026},
}

@misc{JL26b,
  author       = {Jo, Sunghyeon and Lee, Taekyun},
  title        = {Exact universality of online discrepancy},
  howpublished = {Preprint, \href{https://arxiv.org/abs/2610.00103}{arXiv:2610.00103}},
  year         = {2026},
}

@article{ST03,
  author  = {Shcherbina, Mariya and Tirozzi, Brunello},
  title   = {Rigorous solution of the {Gardner} problem},
  journal = {Communications in Mathematical Physics},
  volume  = {234},
  number  = {3},
  pages   = {383--422},
  year    = {2003},
}

@book{Tal11,
  author    = {Talagrand, Michel},
  title     = {Mean field models for spin glasses. {V}olume {II}: Advanced replica-symmetry and low temperature},
  series    = {Ergebnisse der Mathematik und ihrer Grenzgebiete. 3. Folge},
  volume    = {55},
  publisher = {Springer},
  address   = {Heidelberg},
  year      = {2011},
}

@misc{Sto13a,
  author       = {Stojnic, Mihailo},
  title        = {Another look at the {Gardner} problem},
  howpublished = {Preprint, \href{https://arxiv.org/abs/1306.3979}{arXiv:1306.3979}},
  year         = {2013},
}

@book{Mat99,
  author    = {Matou{\v{s}}ek, Ji{\v{r}}{\'\i}},
  title     = {Geometric discrepancy: An illustrated guide},
  series    = {Algorithms and Combinatorics},
  volume    = {18},
  publisher = {Springer-Verlag},
  address   = {Berlin},
  year      = {1999},
}

@misc{AT24,
  author       = {Altschuler, Dylan J. and Tikhomirov, Konstantin},
  title        = {A note on the capacity of the binary perceptron},
  howpublished = {Preprint, \href{https://arxiv.org/abs/2401.15092}{arXiv:2401.15092}},
  year         = {2024},
}

@inproceedings{BNSX22,
  author    = {Bolthausen, Erwin and Nakajima, Shuta and Sun, Nike and Xu, Changji},
  title     = {{Gardner} formula for {Ising} perceptron models at small densities},
  booktitle = {Proceedings of the 35th Conference on Learning Theory (COLT)},
  pages     = {1787--1911},
  year      = {2022},
}

@article{Alt23,
  author  = {Altschuler, Dylan J.},
  title   = {Critical window of the symmetric perceptron},
  journal = {Electronic Journal of Probability},
  volume  = {28},
  pages   = {123},
  year    = {2023},
}

@inproceedings{SS23,
  author    = {Sah, Ashwin and Sawhney, Mehtaab},
  title     = {Distribution of the threshold for the symmetric perceptron},
  booktitle = {2023 IEEE 64th Annual Symposium on Foundations of Computer Science (FOCS)},
  pages     = {2369--2382},
  year      = {2023},
}

@article{KMRTSZ07,
  author  = {Krzakala, Florent and Montanari, Andrea and Ricci-Tersenghi, Federico and Semerjian, Guilhem and Zdeborov{\'a}, Lenka},
  title   = {{Gibbs} states and the set of solutions of random constraint satisfaction problems},
  journal = {Proceedings of the National Academy of Sciences},
  volume  = {104},
  number  = {25},
  pages   = {10318--10323},
  year    = {2007},
}

@article{ZM08b,
  author  = {Zdeborov{\'a}, Lenka and M{\'e}zard, Marc},
  title   = {Constraint satisfaction problems with isolated solutions are hard},
  journal = {Journal of Statistical Mechanics: Theory and Experiment},
  volume  = {2008},
  number  = {12},
  pages   = {P12004},
  year    = {2008},
}

@article{Bar25,
  author  = {Barbier, Damien},
  title   = {Finding the right path: statistical mechanics of connected solutions in constraint satisfaction problems},
  journal = {Journal of Statistical Mechanics: Theory and Experiment},
  volume  = {2026},
  number  = {8},
  pages   = {083401},
  year    = {2026},
}

@misc{Bar26,
  author       = {Barbier, Damien},
  title        = {When low-loss paths make a binary neuron trainable: detecting algorithmic transitions with the connected ensemble},
  howpublished = {Preprint, \href{https://arxiv.org/abs/2601.23241}{arXiv:2601.23241}},
  year         = {2026},
}

@misc{Sto25,
  author       = {Stojnic, Mihailo},
  title        = {Rare dense solutions clusters in asymmetric binary perceptrons -- local entropy via fully lifted {RDT}},
  howpublished = {Preprint, \href{https://arxiv.org/abs/2506.19276}{arXiv:2506.19276}},
  year         = {2025},
}

@article{GS17a,
  author  = {Gamarnik, David and Sudan, Madhu},
  title   = {Limits of local algorithms over sparse random graphs},
  journal = {The Annals of Probability},
  volume  = {45},
  number  = {4},
  pages   = {2353--2376},
  year    = {2017},
}

@article{RV17,
  author  = {Rahman, Mustazee and Vir{\'a}g, B{\'a}lint},
  title   = {Local algorithms for independent sets are half-optimal},
  journal = {The Annals of Probability},
  volume  = {45},
  number  = {3},
  pages   = {1543--1577},
  year    = {2017},
}

@article{Wei22,
  author  = {Wein, Alexander S.},
  title   = {Optimal low-degree hardness of maximum independent set},
  journal = {Mathematical Statistics and Learning},
  volume  = {4},
  number  = {3},
  pages   = {221--251},
  year    = {2022},
}

@inproceedings{BH22,
  author    = {Bresler, Guy and Huang, Brice},
  title     = {The algorithmic phase transition of random {$k$}-{SAT} for low degree polynomials},
  booktitle = {2021 IEEE 62nd Annual Symposium on Foundations of Computer Science (FOCS)},
  pages     = {298--309},
  year      = {2021},
}

@inproceedings{HS25b,
  author    = {Huang, Brice and Sellke, Mark},
  title     = {Tight {Lipschitz} hardness for optimizing mean field spin glasses},
  booktitle = {2022 IEEE 63rd Annual Symposium on Foundations of Computer Science (FOCS)},
  pages     = {312--322},
  year      = {2022},
}

@article{DGH25,
  author  = {Du, Hang and Gong, Shuyang and Huang, Rundong},
  title   = {The algorithmic phase transition of random graph alignment problem},
  journal = {Probability Theory and Related Fields},
  volume  = {191},
  number  = {3--4},
  pages   = {1233--1288},
  year    = {2025},
}

@misc{GKW25,
  author       = {Gamarnik, David and K{\i}z{\i}lda{\u{g}}, Eren C. and Warnke, Lutz},
  title        = {Optimal hardness of online algorithms for large independent sets},
  howpublished = {Preprint, \href{https://arxiv.org/abs/2504.11450}{arXiv:2504.11450}},
  year         = {2025},
}

@article{Tal00,
  author  = {Talagrand, Michel},
  title   = {Intersecting random half-spaces: toward the {Gardner}--{Derrida} formula},
  journal = {The Annals of Probability},
  volume  = {28},
  number  = {2},
  pages   = {725--758},
  year    = {2000},
}

@misc{Nak25,
  author       = {Nakajima, Shuta},
  title        = {Uniqueness of replica-symmetric saddle point for {Ising} perceptron},
  howpublished = {Preprint, \href{https://arxiv.org/abs/2512.23195}{arXiv:2512.23195}},
  year         = {2025},
}

@misc{KL26,
  author       = {K{\i}z{\i}lda{\u{g}}, Eren C. and Li, Shuangping},
  title        = {Stable and online algorithms for random matrix discrepancy},
  howpublished = {Preprint, \href{https://arxiv.org/abs/2610.01591}{arXiv:2610.01591}},
  year         = {2026},
}

@article{KOWZ24,
  author  = {Kunisky, Dmitriy and Oertel, Timm and Wengiel, Nicola and Zhang, Peiyuan},
  title   = {Asymptotic bounds and online algorithms for average-case matrix discrepancy},
  journal = {Random Structures \& Algorithms},
  volume  = {69},
  number  = {3},
  pages   = {e70102},
  year    = {2026},
}

@misc{AT26,
  author       = {Altschuler, Dylan J. and Tikhomirov, Konstantin},
  title        = {The threshold for online balancing of i.i.d.\ binary vectors},
  howpublished = {Preprint, \href{https://arxiv.org/abs/2609.14975}{arXiv:2609.14975}},
  year         = {2026},
}

@misc{Ade26,
  author       = {Aden-Ali, Ishaq},
  title        = {Optimal online discrepancy minimization in linear time},
  howpublished = {Preprint, \href{https://arxiv.org/abs/2607.04388}{arXiv:2607.04388}},
  year         = {2026},
}

@inproceedings{GRS26,
  author    = {Gamarnik, David and R{\'a}cz, Mikl{\'o}s Z. and Schoenbach, Gabe},
  title     = {Optimal hardness of online algorithms for large common induced subgraphs},
  booktitle = {Proceedings of the 39th Conference on Learning Theory (COLT)},
  pages     = {2537--2560},
  year      = {2026},
}

@misc{BBMPRSZ26,
  author       = {Benedetti, Marco and Bogdanov, Andrej and Malatesta, Enrico M. and M{\'e}zard, Marc and Perrupato, Gianmarco and Rosen, Alon and Schwartzbach, Nikolaj I. and Zecchina, Riccardo},
  title        = {Collision resistance of single-layer neural nets},
  howpublished = {Preprint, \href{https://arxiv.org/abs/2606.03807}{arXiv:2606.03807}},
  year         = {2026},
}

@article{Bar24,
  author  = {Barbier, Damien},
  title   = {How to escape atypical regions in the symmetric binary perceptron: a journey through connected-solutions states},
  journal = {SciPost Physics},
  volume  = {18},
  number  = {3},
  pages   = {115},
  year    = {2025},
}

@inproceedings{LS25,
  author    = {Li, Shuangping and Schramm, Tselil},
  title     = {Some easy optimization problems have the overlap-gap property},
  booktitle = {Proceedings of the 38th Conference on Learning Theory (COLT)},
  pages     = {3582--3622},
  year      = {2025},
}

@inproceedings{HS25a,
  author    = {Huang, Brice and Sellke, Mark},
  title     = {Strong low degree hardness for stable local optima in spin glasses},
  booktitle = {2026 IEEE 67th Annual Symposium on Foundations of Computer Science (FOCS)},
  year      = {2026},
}

@inproceedings{Rot17,
  author    = {Rothvoss, Thomas},
  title     = {Constructive discrepancy minimization for convex sets},
  booktitle = {2014 IEEE 55th Annual Symposium on Foundations of Computer Science (FOCS)},
  pages     = {140--145},
  year      = {2014},
}

@article{ES18,
  author  = {Eldan, Ronen and Singh, Mohit},
  title   = {Efficient algorithms for discrepancy minimization in convex sets},
  journal = {Random Structures \& Algorithms},
  volume  = {53},
  number  = {2},
  pages   = {289--307},
  year    = {2018},
}

@inproceedings{AchlioptasRicciTersenghi2006,
  author    = {Achlioptas, Dimitris and Ricci-Tersenghi, Federico},
  title     = {On the solution-space geometry of random constraint satisfaction problems},
  booktitle = {Proceedings of the 38th Annual ACM Symposium on Theory of Computing (STOC)},
  pages     = {130--139},
  year      = {2006},
}

@article{GS17b,
  author  = {Gamarnik, David and Sudan, Madhu},
  title   = {Performance of sequential local algorithms for the random {NAE}-{$K$}-{SAT} problem},
  journal = {SIAM Journal on Computing},
  volume  = {46},
  number  = {2},
  pages   = {590--619},
  year    = {2017},
}

@article{CGPR19,
  author  = {Chen, Wei-Kuo and Gamarnik, David and Panchenko, Dmitry and Rahman, Mustazee},
  title   = {Suboptimality of local algorithms for a class of max-cut problems},
  journal = {The Annals of Probability},
  volume  = {47},
  number  = {3},
  pages   = {1587--1618},
  year    = {2019},
}

@misc{BBMPRSZ25,
  author       = {Benedetti, Marco and Bogdanov, Andrej and Malatesta, Enrico M. and M{\'e}zard, Marc and Perrupato, Gianmarco and Rosen, Alon and Schwartzbach, Nikolaj I. and Zecchina, Riccardo},
  title        = {Overlap gap and computational thresholds in the square wave perceptron},
  howpublished = {Preprint, \href{https://arxiv.org/abs/2506.05197}{arXiv:2506.05197}},
  year         = {2025},
}

\end{document}